\documentclass[a4paper,11pt]{article}
\DeclareUnicodeCharacter{2212}{\ensuremath{-}}
\pdfoutput=1 
\usepackage{jheppub} 
\usepackage[numbers]{natbib}
\usepackage[T1]{fontenc} 
\usepackage{hyperref}
\usepackage{bbold}
\usepackage{float}
\usepackage{mathrsfs}
\usepackage{feynman}
\usepackage{yhmath}
\usepackage{xcolor}
\usepackage{adjustbox}
\usepackage{blindtext}
\usepackage{multirow}
\usepackage{makecell} 
\usepackage{multicol}
\usepackage{caption}
\usepackage{booktabs}
\usepackage{subcaption}
\usepackage{verbatim}
\usepackage{amsmath}
\usepackage{relsize}
\usepackage{siunitx}
\usepackage{bm, nicefrac}
\usepackage[capitalise]{cleveref}
\usepackage[normalem]{ulem}
\usepackage{array}
\usepackage{tabularx}

\newcolumntype{R}[1]{>{\raggedleft\arraybackslash}p{#1}}
\newcolumntype{L}[1]{>{\raggedright\arraybackslash}p{#1}}
\newcolumntype{C}[1]{>{\centering\arraybackslash}p{#1}}

\newcommand{\beq}{\begin{equation}}
\newcommand{\eeq}{\end{equation}}
\newcommand{\beqn}{\begin{eqnarray}}
\newcommand{\eeqn}{\end{eqnarray}}

\newcommand{\GeV}{\ensuremath{\,\text{GeV}}\xspace}

\newcommand{\GF}{\ensuremath{G_\mu}}
\newcommand{\alphas}{\ensuremath{\alpha_\text{s}}\xspace}

\newcommand{\rd}{\mathrm d}

\title{Diboson production in the SMEFT at dimension-8}
\author[a,b,c]{Hesham El Faham,}
\author[a]{Giuseppe Ventura,}
\author[a,d]{Eleni Vryonidou\,}

\affiliation[a]{Department of Physics and Astronomy, University of Manchester, Oxford Road, Manchester M13~9PL, United Kingdom}
\affiliation[b]{TIFLab, Universit\`a degli Studi di Milano, Via Celoria 16, 20133 Milano, Italy}
\affiliation[c]{INFN, Sezione di Milano, Via Celoria 16, 20133 Milano, Italy}
\affiliation[d]{Department of Physics, University of Cyprus, P.O. Box 20537, 1678 Nicosia, Cyprus}

\emailAdd{hesham.elfaham@unimi.it}
\emailAdd{giuseppe.ventura@manchester.ac.uk}
\emailAdd{vryonidou.eleni@ucy.ac.cy}

\preprint{
\begin{flushright}
TIF-UNIMI-2025-21
\end{flushright}
}

\abstract{We present a comprehensive analysis of dimension-8 and dimension-6 effects in fully leptonic $WZ$ and $WW$ production at the Large Hadron Collider (LHC)  within the Standard Model (SM) Effective Field Theory (SMEFT). We focus on dimension-8 operators with maximal energy growth in the quark–(anti)quark-initiated production channel and assess their impact differentially through a variety of observables, including polarisation-sensitive ones. Leveraging existing data from measurements at the LHC, we perform fits to quantify the sensitivity of current and future data to dimension-8 effects and evaluate their interplay with squared dimension-6 contributions. By marginalising over the dimension-8 operators we examine the robustness of a dimension-6 SMEFT analysis in diboson production. We find that dimension-8 effects become subdominant only for new-physics scales above 3~TeV. }


\begin{document}

\maketitle
\flushbottom

\section{Introduction}
\label{sec:intro}
In the absence of direct evidence for new resonances or states, the LHC has transitioned into an era of precision searches for new physics, leveraging highly precise differential measurements to indirectly probe deviations from the Standard Model (SM). The Standard Model Effective Field Theory (SMEFT) provides a systematic, model-agnostic framework for characterising these potential deviations via higher-dimensional operators modifying SM interactions.

Electroweak (EW) diboson production stands out as a particularly promising channel to search for new physics (NP) effects. In the fully leptonic final states, $W^{\pm}Z$ and  $W^{+}W^{-}$—hereafter referred to as $WZ$ and $WW$—production channels offer clean experimental signatures and direct sensitivity to triple gauge couplings (TGCs). The ATLAS and CMS collaborations have performed detailed studies of the $WZ$ channel, examining both polarisation states and spin correlations~\cite{ATLAS:2019bsc,CMS:2021icx,ATLAS:2022oge}, benefiting from high signal purity and the ability to fully reconstruct the final state up to a single neutrino. The anticipated large datasets from Run~3 and the High-Luminosity LHC (HL-LHC) will significantly enhance the precision achievable in differential measurements of these processes and extend the kinematic range of the corresponding observables. 

On the theory side, SM predictions have matured substantially, incorporating next-to-next-to-leading order (NNLO) QCD~\cite{Grazzini:2016ctr,Grazzini:2017ckn} and next-to-leading order (NLO) EW corrections~\cite{Biedermann:2016guo,Biedermann:2016yvs,Biedermann:2017oae, Grazzini:2019jkl}, along with advanced parton-shower (PS) matching and resummation techniques~\cite{Lombardi:2021rvg,Lindert:2022qdd,Chiesa:2020ttl,Grazzini:2015wpa, Kallweit:2020gva,Campbell:2022uzw,Dawson:2016ysj,Campbell:2023cha,Gavardi:2023aco}.
Predictions for intermediate polarised bosons in the pole~\cite{Stuart:1991xk,Aeppli:1993cb,Aeppli:1993rs,Denner:2000bj} or narrow-width approximation~\cite{Richardson:2001df,Uhlemann:2008pm,Artoisenet:2012st} are also known up to (N)NLO in QCD~\cite{Denner:2020bcz,Denner:2020eck,Poncelet:2021jmj} and NLO EW~\cite{Le:2022lrp,Le:2022ppa,Denner:2023ehn,Dao:2023pkl} accuracy, including also NLO matching to PS~\cite{Hoppe:2023uux,Pelliccioli:2023zpd}. 

The $WZ$ channel has been extensively studied. At leading order, the process is exclusively $qq-$initiated, whereas NLO order QCD contributions also involve $q(\bar{q})g$ channels. It is well established~\cite{Rubin:2010xp,Grazzini:2019jkl} that in the SM, the NLO QCD corrections are substantial, predominantly due to hard real-emission configurations, particularly those where a high-$p_T$ electroweak boson recoils against a jet and a softer boson, and the large gluon parton luminosity. The NLO electroweak corrections to the $WZ$ fully leptonic final state are moderate—few percent~\cite{Biedermann:2017oae,Grazzini:2019jkl}—but become increasingly negative at high transverse momenta because of sizeable virtual EW Sudakov logarithms~\cite{Denner:2000jv,Accomando:2004de}. 

The combination of precise theory predictions and of precise highly differential measurements render diboson production a great testing ground for EFT analyses. As such, a plethora of SMEFT analyses traditionally concentrated on dimension-6 operators~\cite{Falkowski:2015jaa,Falkowski:2016cxu,Helset:2017mlf,Baglio:2017bfe,Azatov:2017kzw,Franceschini:2017xkh,Chiesa:2018lcs,Baglio:2018bkm,Liu:2018pkg,Grojean:2018dqj,Azatov:2019xxn,Baglio:2019uty,Baglio:2020oqu,Ellis:2020unq,Degrande:2021zpv,Rossia:2023hen,Degrande:2023iob,Aoude:2023hxv,Degrande:2024bmd,ElFaham:2024uop,Thomas:2024dwd, Haisch:2025jqr}, have explored diboson production. SMEFT analyses of diboson production have explored how to resurrect the typically suppressed interference effects from the dimension-6 triple gauge coupling operator~\cite{Panico:2017frx,Azatov:2017kzw,Franceschini:2017xkh,Azatov:2019xxn}. Recently, Ref.~\cite{Mantani:2025bqu} has emphasised that a linear EFT interpretation of diboson measurements is casting doubts on the validity of the EFT for weakly coupled UV scenarios.
Due to these observations and the particular nature of the diboson process, compelling arguments motivate extending the exploration also to dimension-8 operators. Notably, dimension-8 operators introduce novel Lorentz and gauge structures that are absent at dimension-6, such as the neutral triple gauge coupling (nTGC)—see, for example, Ref.~\cite{Ellis:2022zdw}. Additionally, at $\mathscr{O}(\Lambda^{-4})$, dimension-8 contributions naturally appear alongside the squared dimension-6 ones, making their inclusion, a priori, essential for a consistent EFT expansion. Motivated by these arguments, several studies have appeared in the literature exploring higher dimension effects in diboson production~\cite{Corbett:2023qtg,Degrande:2023iob,Ellis:2023zim,Ellis:2021dfa,Martin:2023tvi,Bellm:2016cks}. Whilst these studies focus on the quark initiated channel, a recent study investigated dimension-8 effects in diboson production, focusing on gluon-initiated \(WW\) production~\cite{Gillies:2024mqp}, and concluded that the impact of dimension-8 SMEFT contributions is negligible. 

In this work, we aim to establish, in a model independent manner, the size of the dimension-8 contributions and to determine whether these contributions can significantly influence the constraints set on the dimension-6 triple gauge interactions and thus invalidate the EFT interpretation of diboson measurements within the dimension-6 SMEFT.  We focus on a set of dimension-8 operators that induce energy-growing contributions scaling as $s^2/\Lambda^4$ in quark–(anti)quark-initiated—referred to as $qq$-initiated—diboson production as classified in Ref.~\cite{Degrande:2023iob}. 

We present a comprehensive analysis of dimension-8 SMEFT effects in $WZ$ and $WW$ production at the LHC at $\sqrt{s} = 13$\,TeV. Our analysis examines a broad spectrum of differential and angular observables, including polarisation-sensitive distributions, within both an inclusive and an experimentally realistic fiducial setups. We also examine the angular coefficients, ones which are directly related to the polarisation fractions, and investigate possible contributions from the SMEFT at both the dimension-6 and dimension-8 levels. Furthermore, we perform fits to existing LHC measurements as well as HL-LHC projections to assess the sensitivity of current and future data to dimension-8 effects, investigating the interplay between linear dimension-8 and quadratic dimension-6 contributions.

This paper is structured as follows:~\cref{sec:theory,sec:comp_setup} outline our theoretical framework and computational methods, respectively. Our predictions and results for $WZ$ and $WW$ production are presented and discussed in~\cref{sec:results}, and a sensitivity study based on current measurements and HL-LHC projections is showcased in~\cref{sec:fit}. We summarise our findings in~\cref{sec:conclusions}.

\section{Theoretical framework}
\label{sec:theory}
We study diboson, $VV$, production at the LHC, incorporating dimension-6 and dimension-8 SMEFT modifications. A typical SMEFT expansion takes the form:
\begin{equation}
    \mathcal{L}_{\rm EFT}
    = \mathcal{L}_{\rm SM}
    + \sum_i \frac{C_i^{(6)}\,\mathcal{O}_i^{(6)}}{\Lambda^2}
    + \sum_j \frac{C_j^{(8)}\,\mathcal{O}_j^{(8)}}{\Lambda^4}
    + \mathscr{O}(\Lambda^{-6})\,.
\label{eq:EFT_Lagrangian}
\end{equation}
In physical observables the squared contributions of dimension-6 operators scale as $\mathscr{O}(\Lambda^{-4})$, and thus appear at the same order as the linear contributions from dimension-8 operators. Our aim is to assess the phenomenological impact of the latter in comparison to the former.

Dimension-6 SMEFT operators have been systematically classified, most notably in the so-called Warsaw basis~\cite{Grzadkowski:2010es}. Bases for dimension-8 operators have also been  constructed~\cite{Li:2020gnx,Murphy:2020rsh} and a generalisation of the basis construction at any mass dimension has been achieved~\cite{Li:2022tec,Harlander:2023psl,Harlander:2023ozs}. A key challenge at this order lies in the sheer number of independent operators that can be constructed~\cite{Murphy:2020rsh,Li:2020gnx}: whilst dimension-8 terms allow for richer Lorentz and gauge structures than their dimension-6 counterparts, they do so at the cost of introducing a significantly larger operator basis. We will now discuss the operators we consider in this work. 

\paragraph{Dimension-6} we focus on the modifications to the TGCs induced by the CP-even, \(\mathcal{O}_{W}\), and CP-odd, \(\mathcal{O}_{\widetilde{W}}\), dimension-6 operators:
\begin{equation}\label{operators}
    \epsilon_{ijk} W_{\mu\nu}^{i} W^{j,\nu\rho} W^{k,\mu}_{\rho}, \hspace{1cm} \epsilon_{ijk} \tilde{W}_{\mu\nu}^{i} W^{j,\nu\rho} W^{k,\mu}_{\rho},
\end{equation}
where \(\tilde{W}_{\mu\nu}\) denotes the dual field strength tensor. Operators modifying \(Vq\bar{q}\) interactions also enter in diboson production but these are sufficiently well constrained and thus do not any affect any constraints set on $C_W$. This result is confirmed by the recent global SMEFT fit in Ref.~\cite{Celada:2024mcf}, which finds nearly identical individual and marginalised bounds on the CP-even coefficient \(C_W\), with the full sensitivity driven by diboson production at the LHC. This justifies our choice of considering only the operators of~\cref{operators} at dimension-6. 

In an analysis of recent diboson measurements~\cite{ElFaham:2024uop}, we have explored how constraints on these operators depend on whether the analysis includes only the linear or both linear and quadratic contributions. Quadratic bounds are significantly more stringent than the linear ones. This is related to the helicity selection rules which suppress the interference of the dimension-6 operator with the SM. Quadratic dominated bounds often cast doubt on the validity of the EFT and motivates the inclusion of all effects at \(\mathscr{O}(\Lambda^{-4})\).

At \(\mathscr{O}(\Lambda^{-4})\), one encounters contributions from dimension-8 operators at linear order, the squared terms of dimension-6 operators, as well as their double insertions. In addition to the operators of~\cref{operators}, operators introducing modifications to the \(Vq\bar{q}\) couplings are, in principle, contributing—either via double insertions or through the quadratic contributions from single insertions—we have verified that both types of modifications exhibit suppressed energy growth relative to the linear contributions from dimension-8 operators. They are therefore expected to be subdominant. It is worth noting that we refer here to modifications of the \(Vq\bar{q}\) vertex induced by current-type SMEFT operators; modifications arising from dipole-type SMEFT operators are suppressed under Minimal Flavour Violation assumptions~\cite{DAmbrosio:2002vsn}. Finally, possible deformations in the lepton sector that could enter via EW boson decays are also omitted as they are already tightly constrained by LEP data~\cite{Celada:2024mcf}. 

In summary, our analysis includes the purely dimension-6 modification arising from $\mathcal{O}_{W}$ and $\mathcal{O}_{\tilde{W}}$, as well as the maximally growing dimension-8 operators discussed below. 

\paragraph{Dimension-8} we make use of the classification of dimension-8 operators relevant to $VV$ production in the SMEFT as presented in Ref.~\cite{Degrande:2023iob} where the authors have listed all dimension-8 operators that generate contact interactions for the $qq$-initiated production channels of $WZ$ and $WW$ inducing an $s^2/\Lambda^4$ enhancement in the amplitude in the high energy regime. These are listed in~\cref{tab:qq_operators}.
\begin{table}[ht]
    \centering
    \renewcommand{\arraystretch}{1.3}
    \centering
\begin{tabular}{l|l}
\hline
$\mathcal{O}_1 = iB^{\mu }{}{}_{\nu } B^{\nu }{}{}_{\lambda } (\bar{d}\gamma^\lambda  \overleftrightarrow{D}_\mu d)$ & $\mathcal{O}_{10}=i W^{I}{}^{\mu }{}{}_{\nu } W^{I}{}^{\nu }{}{}_{\lambda } \left(\bar{q}\gamma^{\lambda } \overleftrightarrow{D}_{\mu } q^{}\right)$ \\
$\mathcal{O}_2 = iB^{\mu }{}{}_{\nu } B^{\nu }{}{}_{\lambda } (\bar{u}\gamma^\lambda  \overleftrightarrow{D}_\mu u)$ & $\mathcal{O}_{11}=i \epsilon ^{IJK} W^{I}{}^{\mu }{}{}_{\nu } W^{J}{}^{\nu }{}{}_{\lambda } \left(\bar{q}^{i} \gamma^{\lambda } \tau ^{Kj}_i\overleftrightarrow{D}_{\mu } q_j\right)$ \\
$\mathcal{O}_3 = iB^{\mu }{}{}_{\nu } B^{\nu }{}{}_{\lambda } \left(\bar{q} \gamma^{\lambda } \overleftrightarrow{D}_{\mu } q\right)$ & $\mathcal{O}_{12}=i \epsilon ^{IJK} \tilde{W}^{I}{}^{\mu }{}{}_{\nu } W^{J}{}^{\nu }{}{}_{\lambda } \left(\bar{q}^{i} \gamma^{\lambda }\tau ^{Kj}_i\overleftrightarrow{D}_{\mu } q_{j}\right)$ \\
$\mathcal{O}_4 = i W^{I\mu}{}_{\lambda}B^{\nu\lambda} \left(\bar{q}^{i} \gamma_{\nu } \tau ^{Ij}_i\overleftrightarrow{D}_{\mu } q_{j}\right)$ & $\mathcal{O}_{13}=i\epsilon ^{IJK} W^{I}{}^{\mu }{}{}_{\nu } \tilde{W}^{J}{}^{\nu }{}{}_{\lambda } \left(\bar{q}^{i} \gamma^{\lambda } \tau ^{Kj}_i\overleftrightarrow{D}_{\mu } q_{j}\right)$ \\
$\mathcal{O}_5 = i W^{I\mu}{}_{\lambda}\tilde{B}^{\nu\lambda} \left(\bar{q}^{i} \gamma_{\nu } \tau ^{Ij}_i\overleftrightarrow{D}_{\mu } q_{j}\right)$ & $\mathcal{O}_{14}=i \left(\bar{u} \gamma^{\lambda } \overleftrightarrow{D}_{\mu } u^{}\right)\left(D_\lambda H^\dagger  D^{\mu}H\right)$ \\
$\mathcal{O}_6 = i W^{I\nu}{}_{\lambda}B^{\mu\lambda} \left(\bar{q}^{i} \gamma_{\nu } \tau ^{Ij}_i\overleftrightarrow{D}_{\mu } q_{j}\right)$ & $\mathcal{O}_{15}=i \left(\bar{d} \gamma^{\lambda } \overleftrightarrow{D}_{\mu } d^{}\right)\left(D_\lambda H^\dagger  D^{\mu}H\right)$ \\
$\mathcal{O}_7 = i W^{I\nu}{}_{\lambda}\tilde{B}^{\mu\lambda} \left(\bar{q}^{i} \gamma_{\nu } \tau ^{Ij}_i\overleftrightarrow{D}_{\mu } q_{j}\right)$ & $\mathcal{O}_{16}=i \left(\bar{q} \gamma^{\lambda } \overleftrightarrow{D}_{\mu } q^{}\right)\left(D_\lambda H^\dagger  D^{\mu}H\right)$ \\
$\mathcal{O}_8 = iW^{I}{}^{\mu }{}{}_{\nu } W^{I}{}^{\nu }{}{}_{\lambda }(\bar{d}\gamma^\lambda  \overleftrightarrow{D}_\mu d)$ & $\mathcal{O}_{17}=i \left(\bar{q} \gamma^{\lambda }\tau ^K\overleftrightarrow{D}_{\mu } q\right)\left(D_\lambda H^\dagger \tau^K D^{\mu}H\right)$ \\
$\mathcal{O}_9 = iW^{I}{}^{\mu }{}{}_{\nu } W^{I}{}^{\nu }{}{}_{\lambda }(\bar{u}\gamma^\lambda  \overleftrightarrow{D}_\mu u)$ & $\mathcal{O}_{18}=i(\bar{u}\gamma^\mu  \overleftrightarrow{D}^\nu {d})\epsilon^{ij}(D^\mu H_{i} D^\nu H_{j})$ \\ [0.0ex]
\hline
\end{tabular}
    \caption{List of energy-growing dimension-8 operators contributing to $qq$-initiated $VV$ production as listed in Ref.~\cite{Degrande:2023iob}. $W_{\mu\nu}^I$ and $B_{\mu\nu}$ are the $\mathrm{SU}(2)_{L}$ and $\mathrm{U}(1)_{Y}$ field-strength tensors, $H$ is the Higgs doublet, $q$ indicates the $\mathrm{SU}(2)_{L}$ symmetric quark doublet, and $u$ and $d$ are the right-handed up and down singlet quark fields, respectively. Generation indices for the quark fields are understood.
    }
\label{tab:qq_operators}
\end{table}
It is worth noting that not all operators listed there are relevant for the processes considered in this work, namely $WZ$ and $WW$. In particular, some operators contribute exclusively to $ZZ$ production and are therefore omitted from our analysis. The subsets of operators relevant to our study are the following:
\begin{equation}
\begin{aligned}
    WZ \, : & \quad \{ \mathcal{O}_4, \mathcal{O}_5, \mathcal{O}_6, \mathcal{O}_7, \mathcal{O}_{11}, \mathcal{O}_{12}, \mathcal{O}_{13}, \mathcal{O}_{17}, \mathcal{O}_{18} \},\\
    WW \, : & \quad \{ \mathcal{O}_8, \mathcal{O}_9, \mathcal{O}_{10}, \mathcal{O}_{11}, \mathcal{O}_{12}, \mathcal{O}_{13}, \mathcal{O}_{14}, \mathcal{O}_{15}, \mathcal{O}_{16}, \mathcal{O}_{17} \},
\end{aligned}
\label{eq:rel_ops}
\end{equation}
where only three operators, $\mathcal{O}_5$, $\mathcal{O}_7$, and $\mathcal{O}_{11}$, are $\rm CP$-odd, whilst all others are $\rm CP$-even.

Among the relatively large number of dimension-8 operators contributing to $WZ$ and $WW$ production via the $qq$-initiated channel presented in~\cref{eq:rel_ops}, a hierarchy emerges in their energy growth either at the amplitude level or after integration over phase space. For completeness and to motivate our focus on the leading energy-growth effects, we summarise here the key arguments of Ref.~\cite{Degrande:2023iob} regarding the scaling behaviour of the operators in~\cref{eq:rel_ops}. These considerations explain the origin of the hierarchy—both at the amplitude level and following phase-space integration—and justify restricting our analysis to the subset of operators that dominate the energy growth.

\paragraph{{$\mathbf{WW}$} at dimension-8} since the SM diboson amplitude vanishes for opposite-helicity states, any non-zero interference between the EFT and SM amplitudes requires a helicity flip. Such flips can occur, for example, through insertions of the gauge-boson mass, \(m_V\), which introduce a suppression factor of at least one power of mass.
Consequently, the resulting interference amplitude from \(\mathcal{O}_{(8,9)}\) grows only as $\sim s$ at high energy. In contrast, the operators \(\mathcal{O}_{(14,15)}\) probe the longitudinal modes of the gauge bosons.  Owing to their additional derivative structure, their contributions to the amplitude scale as $\sim s^{2}$ at high energy. 

The operator \(\mathcal{O}_{(11)}\) is CP-odd, resulting in a vanishing interference with the CP-even SM amplitude for the $2\to2$ amplitudes. Interference involving CP-odd operators can re-emerge in \(2 \to 4\) amplitudes~\cite{Azatov:2017kzw,Azatov:2019xxn,Degrande:2021zpv,Degrande:2023iob}. However, we find that such contributions are generally negligible in our analysis, particularly in the \(WW\) channel, where they are statistically compatible with zero—similar observations have been found in Ref.~\cite{ElFaham:2024uop}. For this reason, we do not present CP-odd results for the \(WW\) process.

This leaves the operators \(\mathcal{O}_{(16,17,10,12,13)}\), which each yield amplitudes scaling as $\sim s^{2}$ at high energy. However, after the angular phase‐space integration, only \(\mathcal{O}_{(10,12,13)}\) retain this leading growth. The operators \(\mathcal{O}_{(8,9,11,14,15)}\) integrate to zero, whilst \(\mathcal{O}_{(16,17)}\) suffer a \(1/s\) suppression. This hierarchy is most transparently understood via the \(J\)-basis formalism and partial wave decomposition, as detailed in Ref.~\cite{Degrande:2023iob}.

\paragraph{{$\mathbf{WZ}$} at dimension-8} the operators \(\mathcal{O}_{(5,7,11)}\) are CP-odd. However, unlike in the $WW$ case, possible $2\to2$ non-interference with the SM can be resurrected by including the finite width effects. The operator \(\mathcal{O}_{(18)}\) couples only to right-handed quarks, which also leads to zero interference with the SM. Therefore, the only CP-even operators we need to consider in our study are \(\mathcal{O}_{(4,6,12,13)}\). 

After accounting for the high-energy scaling of the dimension-8 operators, the relevant subsets in~\cref{eq:rel_ops} reduce to smaller sets which we consider in this study, namely
\begin{equation}
  \label{eq:considered_ops}
  WZ: \;\{\,\mathcal{O}_{4},\,\mathcal{O}_{5},\,\mathcal{O}_{6},\,\mathcal{O}_{7},\,\mathcal{O}_{11},\,\mathcal{O}_{12},\,\mathcal{O}_{13}\,\},
  \qquad
   WW: \;\{\,\mathcal{O}_{10},\,\mathcal{O}_{12},\,\mathcal{O}_{13}\,\}.
\end{equation}

\section{Computational setup}
\label{sec:comp_setup} 
We study the production of $WZ$ and $WW$ at the LHC at $\sqrt{s} = 13$\,TeV, focusing on fully leptonic final states, namely
\beq
p p \to e^+ \nu_{e} \hspace{0.1cm} \mu^+\mu^-\,,\qquad 
p p \to e^+ \nu_{e} \hspace{0.1cm} \mu^-\bar{\nu}_\mu \,,
\eeq
for $WZ$ and $WW$, respectively.
Predictions are obtained at NLO in QCD for the SM and when including the dimension-6 SMEFT operators, $\mathcal{O}_W$ and $\mathcal{O}_{\tilde{W}}$, and at LO in the presence of the dimension-8 operators listed in~\cref{eq:considered_ops}.

The calculations are performed in the $G_\mu$ scheme~\cite{Denner:2000bj,Dittmaier:2001ay}, which is well-suited for SMEFT analyses in light of recent recommendations from the LHC EFT Working Group~\cite{Brivio:2021yjb}. The Fermi constant is set to $\GF = 1.16638 \cdot 10^{-5}\,\GeV^{-2}$. Unstable particles are treated using the complex-mass scheme~\cite{Denner:2005fg,Denner:2006ic,Denner:2019vbn}. The masses of the EW gauge bosons are $m_W = 80.379 \GeV$ and $m_Z = 91.1876 \GeV$. The top-quark pole mass and width, relevant for bottom-induced $WW$ production, are $m_t = 172 \GeV$ and $\Gamma_t = 1.47 \GeV$, respectively.

All computations are performed in the five-flavour scheme. Parton distributions are taken from the NNPDF3.1 NLO set~\cite{Ball:2017nwa} with $\alphas(m_{Z}) = 0.118$ accessed via the LHAPDF interface~\cite{Buckley:2014ana}. We adopt fixed renormalisation and factorisation scales, in line with recent SM polarisation analyses~\cite{Denner:2020bcz,Denner:2020eck,Poncelet:2021jmj,Le:2022lrp,Le:2022ppa,Denner:2023ehn,Dao:2023pkl,Hoppe:2023uux,Pelliccioli:2023zpd}, namely
\begin{equation}
    \label{eq:fixedscale}
    \mu_{\rm R} = \mu_{\rm F} = 
    \begin{cases}
    \displaystyle \frac{m_{W} + m_{Z}}{2}, & \text{for } WZ, \\
    m_{W}, & \text{for } WW.
    \end{cases}
\end{equation}
It has been shown in Ref.~\cite{ElFaham:2024uop} that using a dynamical scale yields results consistent with the fixed-scale choice, both at the inclusive level and for differential observables.

The SM and dimension-6 SMEFT predictions for both processes have been cross-validated against the results in Ref.~\cite{ElFaham:2024uop}. SMEFT predictions are obtained using the \texttt{SMEFT@NLO} model~\cite{Degrande:2020evl} implemented in \texttt{MadGraph5\_aMC@NLO}~\cite{Alwall:2014hca}. The model has been extended to include the $\rm{CP}$-odd triple-gauge operator, $\mathcal{O}_{\tilde{W}}$, relevant for our study. For the dimension-8 results, we generated our UFO model~\cite{Degrande:2011ua} based on the \texttt{FeynRules}~\cite{Alloul:2013bka} implementation of the SMEFT developed in~\cite{Degrande:2023iob}. This UFO model was interfaced with \texttt{MadGraph5\_aMC@NLO} to generate parton-level predictions for the relevant processes.

\paragraph{$\mathbf{WZ}$} we consider both inclusive and fiducial selection criteria for $WZ$ production. The inclusive setup imposes only a single requirement: a mass window on the same-flavour, opposite-sign lepton pair, namely
\begin{equation}
    81\GeV < m_{\mu\mu} < 101\GeV\,.
    \label{eq:mllcut}
\end{equation}

The fiducial selection follows closely the event-level criteria adopted in Refs.~\cite{ATLAS:2019bsc,ATLAS:2022oge}, and—in addition to the invariant mass window in~\cref{eq:mllcut}—is defined as:
\begin{align}
 & p_{T}^{e} > 20\GeV\,, \qquad p_{T}^{\mu^{\pm}} > 15\GeV\,, \nonumber\\
 & m_{T}^{W} > 30\GeV\,, \qquad |y_l| < 2.5\,, \nonumber\\
 & \Delta R_{\mu\mu} > 0.2\,, \qquad \Delta R_{\mu^{\pm} e} > 0.3\,, \label{eq:fidSetup}
\end{align}
with the transverse mass of the $W$ boson defined as
\begin{equation}
    m_{T}^{W} = \sqrt{2\,p_{T}^{e}\,p_{T}^{\rm miss} \left(1 - \cos\Delta\Phi_{e, \rm miss}\right)}.
\end{equation}
No veto is applied on additional hadronic activity in the event. Owing to the presence of a single neutrino in the final state, the full kinematics can be reconstructed using an on-shell constraint on the $W$ boson. For this purpose, we adopt the reconstruction method employed in Ref.~\cite{ATLAS:2019bsc}.

\paragraph{$\mathbf{WW}$} we examine three different event selection strategies for the $WW$ process. The first is a fully inclusive setup, imposing no kinematic cuts on the final-state leptons and applying no vetoes on jets. The second setup retains the inclusivity in lepton kinematics but incorporates jet vetoes following the ATLAS analysis in Ref.~\cite{ATLAS:2019rob}, designed to suppress contributions from top-quark and multijet QCD backgrounds through vetoing:
\begin{itemize}
    \item $b$-tagged jets with $p_{T}^{b} > 20\GeV$ and $|\eta_{b}| < 2.5$\,,
    \item light jets with $p_{T}^{j} > 35\GeV$ and $|\eta_{j}| < 4.5$\,.
\end{itemize}
The third configuration implements the full set of fiducial cuts employed in Ref.~\cite{ATLAS:2019rob}, combining the above jet vetoes with additional lepton-level requirements:
\begin{align}
 & p_{T}^{l} > 27\GeV\,,\quad p_{T}^{\rm miss} > 20\GeV\,,\quad p_{T}^{e\mu} > 30\GeV\,, \nonumber\\ 
 & |y_{\mu}| < 2.5\,,\qquad |y_{e}| < 2.47\,,\qquad m_{e\mu} > 55\GeV\,.
\end{align}

\section{Results}
\label{sec:results}
In~\cref{tab:wz_total_xsec_cp_even,tab:wz_total_xsec_cp_odd}, we present inclusive and fiducial cross sections for the SM, together with the relevant dimension-6 and dimension-8 operators in the \(WZ\) channel for CP-even and CP-odd operators, respectively. The linear and the quadratic contributions of the dimension-6 CP-even operator are denoted as $\mathcal{O}_{W}^{\rm int}$ and $\mathcal{O}_{W}^{\rm sq}$, respectively. The reported cross sections assume \(C_{i}=1\) and \(\Lambda=1~\text{TeV}\).
\newcommand{\scalepm}[2]{\ensuremath{^{+#1\%}_{-#2\%}}}
\begin{table}[ht]
\resizebox{\textwidth}{!}{
    \renewcommand{\arraystretch}{1.25}
    \centering
    \begin{tabular}{c|c|c|c|c|c|c|c}
          \hline $\sigma$ [fb] & SM & $\mathcal{O}_{W}^{\rm int}$ & $\mathcal{O}_{W}^{\rm sq}$ & $\mathcal{O}_{4}$ & $\mathcal{O}_{6}$ & $\mathcal{O}_{12}$ &  $\mathcal{O}_{13}$
          \\  \hline 
          incl. 
          & \(98.37(2)\scalepm{5.3}{5.1}\)
          & \(-1.417(1)\scalepm{26}{22}\) 
          & \(12.43(1)\scalepm{0.9}{1.1}\)
          & \(-0.0868(1)\scalepm{5.5}{4.8}\) 
          & \(-0.0891(3)\scalepm{5.8}{5.0}\) 
          & \(-1.613(1)\scalepm{5.9}{5.2}\)
          & \(1.512(1)\scalepm{6.3}{5.5}\)
          \\ 
          fid. 
          & \(35.34(1)\scalepm{5.6}{5.6}\)
          & \(-0.995(2)\scalepm{16}{14}\) 
          & \(6.575(4)\scalepm{1.6}{1.6}\)
          & \(-0.0382(1)\scalepm{5.8}{5.1}\) 
          & \(-0.0353(2)\scalepm{5.0}{4.5}\)
          & \(-0.644(1)\scalepm{6.1}{5.3}\) 
          & \(0.702(1)\scalepm{6.5}{5.6}\)
          \\ \hline 
    \end{tabular}
    }
     \caption{Inclusive (incl.) and fiducial (fid.) $WZ$ cross sections. Shown are the NLO QCD predictions for the SM and for the dimension-6 CP-even operator $\mathcal{O}_{W}$—interference (int) and quadratic (sq) contributions, as well as the contributions of the relevant dimension-8 CP-even operators computed at LO. Numbers in parentheses denote MC statistical uncertainties, whilst percentage values correspond to QCD scale variations.}
    \label{tab:wz_total_xsec_cp_even}
\end{table}
\begin{table}[ht]
\resizebox{\textwidth}{!}{
    \renewcommand{\arraystretch}{1.25}
    \centering
    \begin{tabular}{c|c|c|c|c|c}
          \hline $\sigma$ [fb] & $\mathcal{O}_{\tilde{W}}^{\rm int}$ & $\mathcal{O}_{\tilde{W}}^{\rm sq}$ & $\mathcal{O}_{5}$ & $\mathcal{O}_{7}$ & $\mathcal{O}_{11}$
          \\  \hline 
          incl. 
          & \(-0.182(3)\scalepm{3.6}{4.1}\)
          & \(12.78(1)\scalepm{0.8}{1.1}\) 
          & \(0.0051(1)\scalepm{6.0}{5.2}\)
          & \(0.0044(1)\scalepm{6.7}{5.8}\) 
          & \(-0.0035(1)\scalepm{2.3}{3.2}\) 
          \\ 
          fid. 
          & \(-0.059(1)\scalepm{0.6}{1.6}\)
          & \(6.71(1)\scalepm{1.5}{1.5}\) 
          & \(0.0024(1)\scalepm{6.3}{5.4}\)
          &  \(0.0018(1)\scalepm{6.6}{5.7}\) 
          & \(-0.0018(1)\scalepm{1.8}{1.2}\)
          \\ \hline 
    \end{tabular}
    }
     \caption{Same as~\cref{tab:wz_total_xsec_cp_even} but for CP-odd operators.}
    \label{tab:wz_total_xsec_cp_odd}
\end{table}

At the inclusive level, we find the dimension-8 contributions to be subdominant with respect to the squared contributions of the dimension-6 operator, \(\mathcal{O}_{W}^{\text{sq}}\), both in the inclusive and fiducial setups of the \(WZ\) channel. These conclusions hold for both the CP-even and CP-odd coefficients, as reported in~\cref{tab:wz_total_xsec_cp_even,tab:wz_total_xsec_cp_odd}, respectively. 

For the \(WW\) channel, \cref{tab:ww_total_xsec_cp_even} reports the inclusive cross section in the three setups defined in~\cref{sec:comp_setup}, for the SM as well as for the dimension-6 and dimension-8 CP-even operators. 
\begin{table}[ht]
\resizebox{\textwidth}{!}{
    \renewcommand{\arraystretch}{1.25}
    \centering
    \begin{tabular}{c|c|c|c|c|c|c}
          \hline $\sigma$ [fb] & SM & $\mathcal{O}_{W}^{\rm int}$ & $\mathcal{O}_{W}^{\rm sq}$ & $\mathcal{O}_{10}$ & $\mathcal{O}_{12}$ & $\mathcal{O}_{13}$
          \\  \hline 
          incl. 
          & \(1895.1(5)\scalepm{11}{11}\)
          & \(-8.57(6)\scalepm{17}{15}\)
          &\(40.78(9)\scalepm{1.3}{1.4}\) 
          & \(21.27(1)\scalepm{4.6}{4.2}\)
          &\(-5.10(1)\scalepm{5.1}{4.6}\)
          & \(4.86(1)\scalepm{5.2}{4.7}\) 
          \\ 
          fid. 
          & \(186.8(2)\scalepm{3.6}{4.8}\)
          & \(-0.74(5)\scalepm{14}{12}\)
          & \(10.4(1)\scalepm{27}{39}\) 
          & \(12.06(1)\scalepm{6.7}{5.8}\)
          &  \(-2.39(1)\scalepm{7.4}{6.3}\) 
          &\(2.36(1)\scalepm{7.5}{6.4}\)
          \\ 
          jet-veto-only 
          & \(1004.5(3)\scalepm{3.5}{4.4}\)
          & \(-0.43(2)\scalepm{76}{66}\)
          & \(17.8(2)\scalepm{19}{26}\) 
          & \(21.27(1)\scalepm{4.6}{4.2}\)
          &  \(-5.10(1)\scalepm{5.1}{4.6}\) 
          & \(4.86(1)\scalepm{5.2}{4.7}\)
          \\\hline
    \end{tabular}
    }
     \caption{Same as~\cref{tab:wz_total_xsec_cp_even} but for the $WW$ process. Jet-veto–only cross sections are, as expected, identical at LO and differ only at NLO; hence differences appear only in the first three columns.}
    \label{tab:ww_total_xsec_cp_even}
\end{table}

Overall the dimension-8 contributions are comparatively larger than for $WZ$. We observe that the contributions from \(\mathcal{O}_{12}\) and \(\mathcal{O}_{13}\) remain subdominant compared to the squared dimension-6 terms across all setups, whereas the effects of \(\mathcal{O}_{10}\) are competitive—particularly in the fiducial and jet-veto-only cases. Further discussion of this behaviour will be provided in the following sections.

\subsection{Differential distributions}
In this section, we present differential predictions for a selected set of observables:
\begin{equation}
    \{ m_{4l},\, m_{ij},\, p_T^X,\, \phi^{*}_{i}, \, \cos\theta^{*}_{i},\, \Delta\Phi_{i,j},\, \Delta y_{i,j} \},
\end{equation}
which correspond, respectively, to the invariant mass of the four-lepton system, the invariant mass of particles \( i \) and \( j \), the transverse momentum of particle \( X \), the azimuthal angle and polar angle cosine of particle \( i \), and the azimuthal and rapidity separations between particles \( i \) and \( j \).

The polar and azimuthal angles of a selected decay lepton are computed in the rest frame of its parent EW boson, relative to the boson's flight direction in the diboson centre-of-mass (CM) frame. This frame is widely regarded as the natural choice for diboson studies~\cite{Denner:2020eck,Le:2022lrp,Le:2022ppa,Denner:2023ehn,Dao:2023pkl,Hoppe:2023uux,Pelliccioli:2023zpd}, as it directly accesses polarisation coefficients defined in the diboson CM frame—commonly adopted in experimental analyses. To fully specify the coordinate system~\cite{Boudjema:2009fz}, we set the reference/zero axis for the azimuthal angle by aligning it with the decay plane of the $Z$ boson in the diboson CM frame. 

Throughout this work, the superscript `\({\mathrm{reco.}}\)' denotes observables computed with reconstructed information on the missing transverse momentum, \(p_T^{\text{miss}}\), whereas the absence of this superscript indicates that truth-level information is used. The treatment of polarisation fractions is more involved in this regard, and we comment on this below. 

\begin{figure}[ht]
    \centering
    \begin{subfigure}[b]{0.49\linewidth}
        \centering
       \includegraphics[page=1, width=\textwidth]{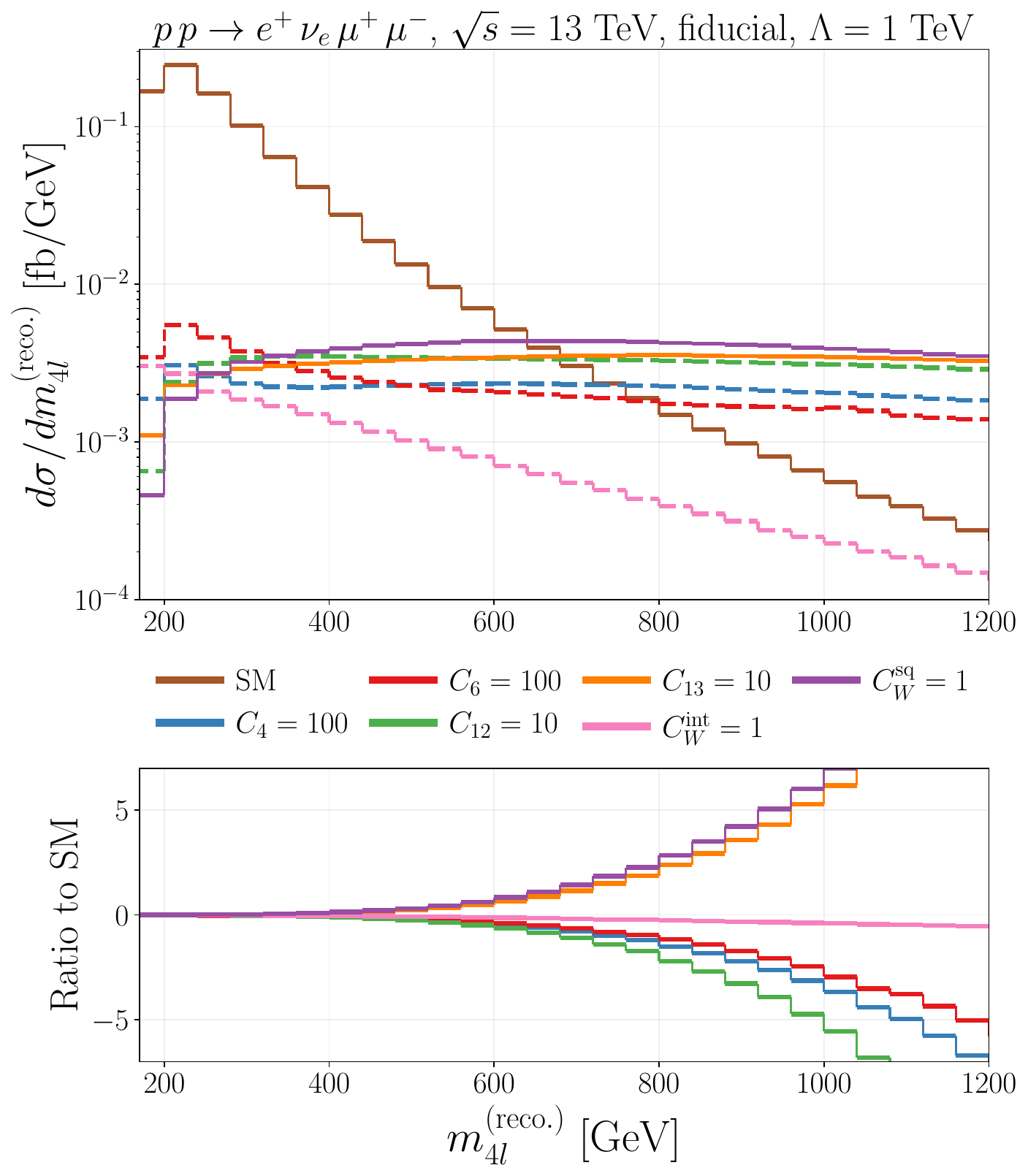}
    \end{subfigure}
    \hfill
    \begin{subfigure}[b]{0.49\linewidth}
        \centering
        \includegraphics[page=2, width=\textwidth]{WZ_plots_even/plots_WZ_even_reconstructed_fiducial.pdf}
    \end{subfigure}
    \caption{Distributions of the invariant mass of the four-lepton system, $m_{4l}$, (left) and the transverse momentum of the $Z$ boson, $p_T^{Z}$, (right) for $WZ$ production in the ATLAS fiducial setup. Predictions shown include the SM as well as linear and quadratic contributions of the dimension-6 SMEFT operator $\mathcal{O}_W$ computed at NLO in QCD, along with the linear dimension-8 contributions ($\mathcal{O}_{4}, \mathcal{O}_{6}, \mathcal{O}_{12}, \mathcal{O}_{13}$) at $\mathscr{O}(\Lambda^{-4})$ computed at LO. The inset displays the ratio of the EFT contributions to the SM. Some coefficients are scaled for visibility. Dashed lines indicate negative contributions.}
    \label{fig:wz_fid_m4l_ptz}
\end{figure}

\begin{figure}[ht]
    \centering
    \begin{subfigure}[b]{0.49\linewidth}
        \centering
        \includegraphics[page=3, width=\textwidth]{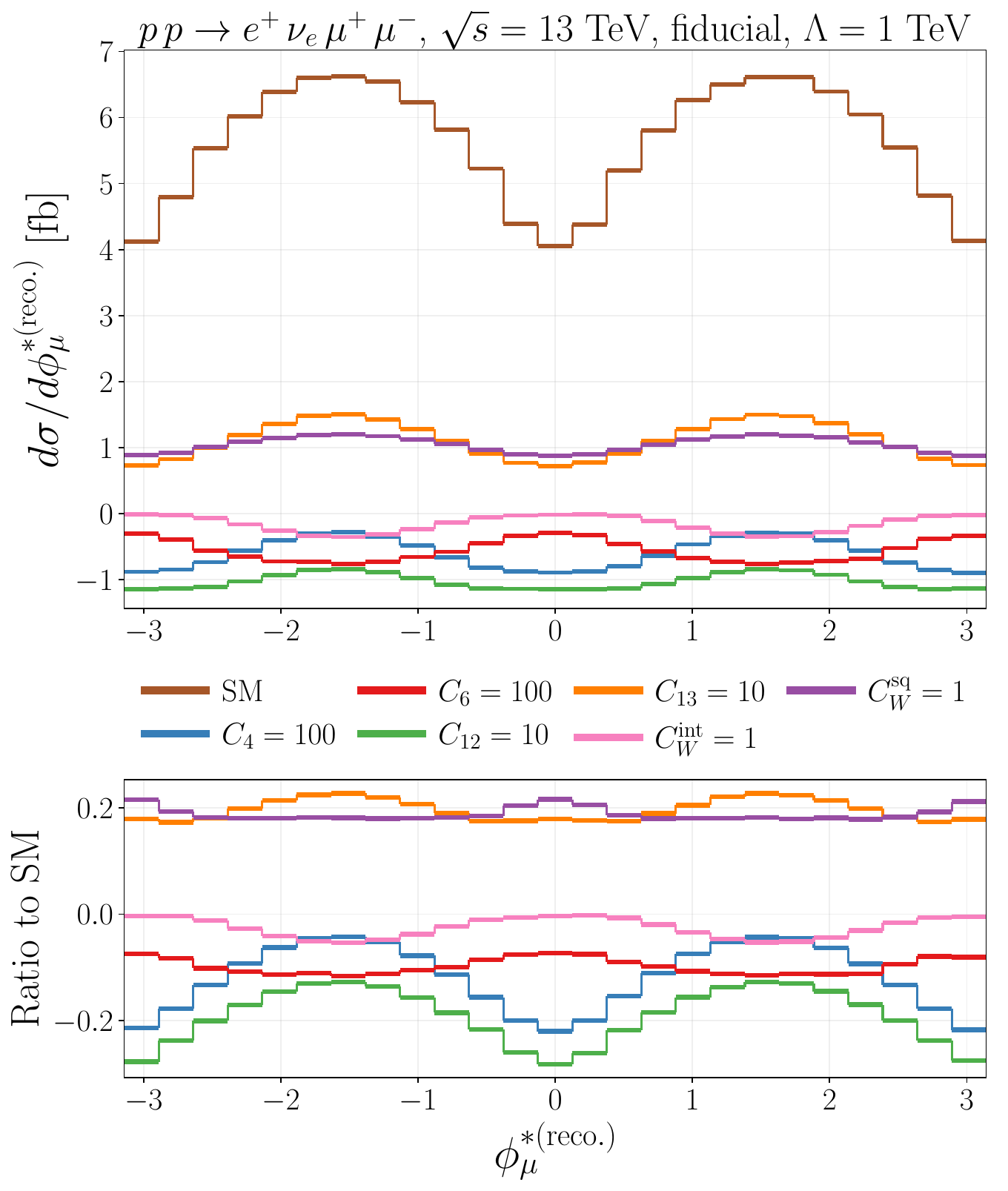}
    \end{subfigure}
    \hfill 
    \begin{subfigure}[b]{0.49\linewidth}
        \centering
        \includegraphics[page=4, width=\textwidth]{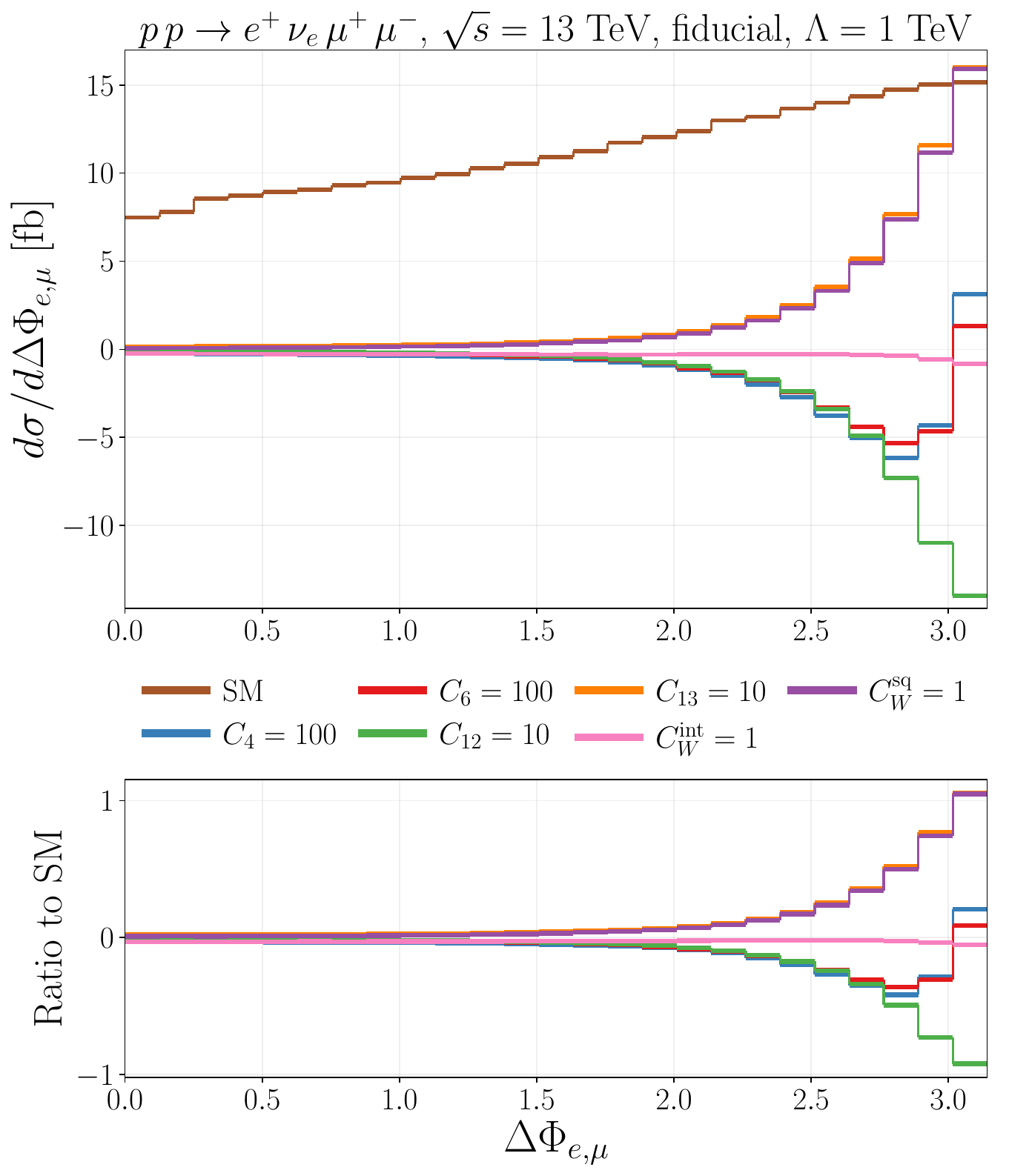}
    \end{subfigure}
    \caption{Same as~\cref{fig:wz_fid_m4l_ptz} but for the azimuthal angle of $\mu$ (left), and the azimuthal separation between $e$ and $\mu$ (right).}
    \label{fig:wz_fid_phi_dphi}
\end{figure}

\begin{figure}[ht]
    \centering
    \begin{subfigure}[b]{0.49\linewidth}
        \centering
       \includegraphics[page=5, width=\textwidth]{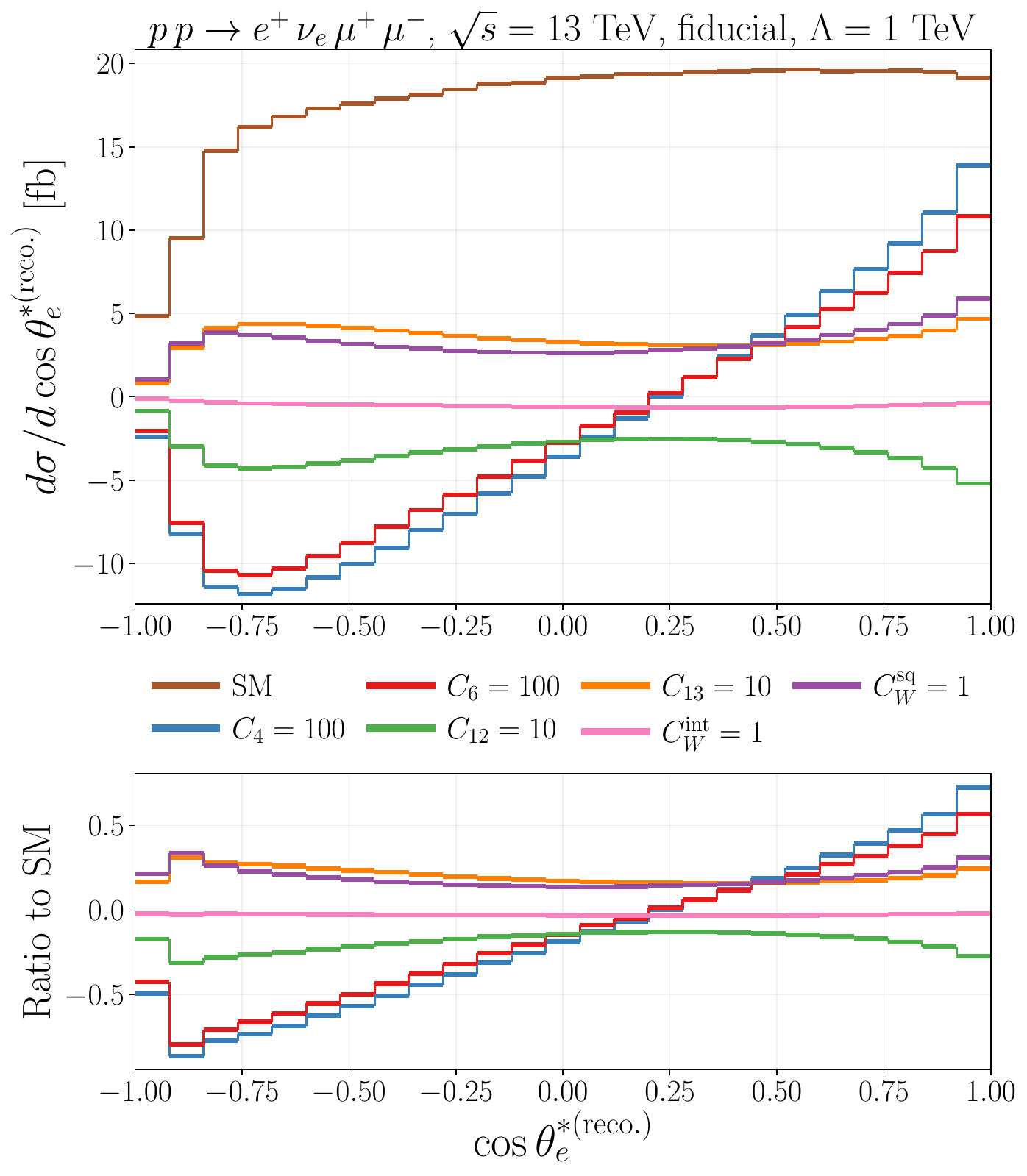}
    \end{subfigure}
    \hfill 
    \begin{subfigure}[b]{0.49\linewidth}
        \centering
       \includegraphics[page=6, width=\textwidth]{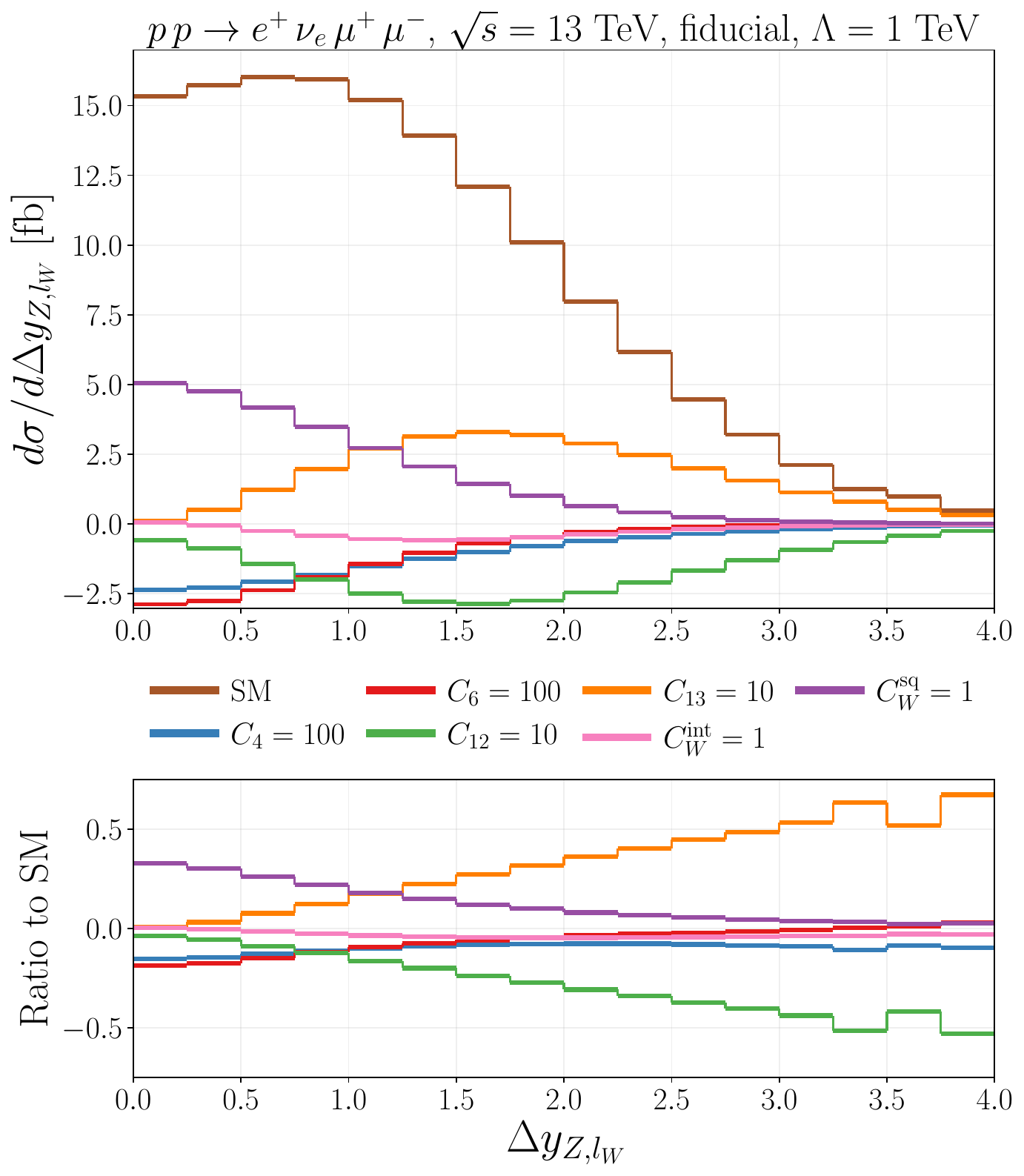}
    \end{subfigure}
    \caption{Same as~\cref{fig:wz_fid_m4l_ptz} but for the polar angle of $e$ (left), and the rapidity separation between $Z$ and the lepton $l$ from the $W$ decay, $l_{W}$, (right).}
    \label{fig:wz_fid_cos_dy}
\end{figure}

\begin{figure}[ht]
    \centering
    \begin{subfigure}[b]{0.49\linewidth}
        \centering
       \includegraphics[page=8, width=\textwidth]{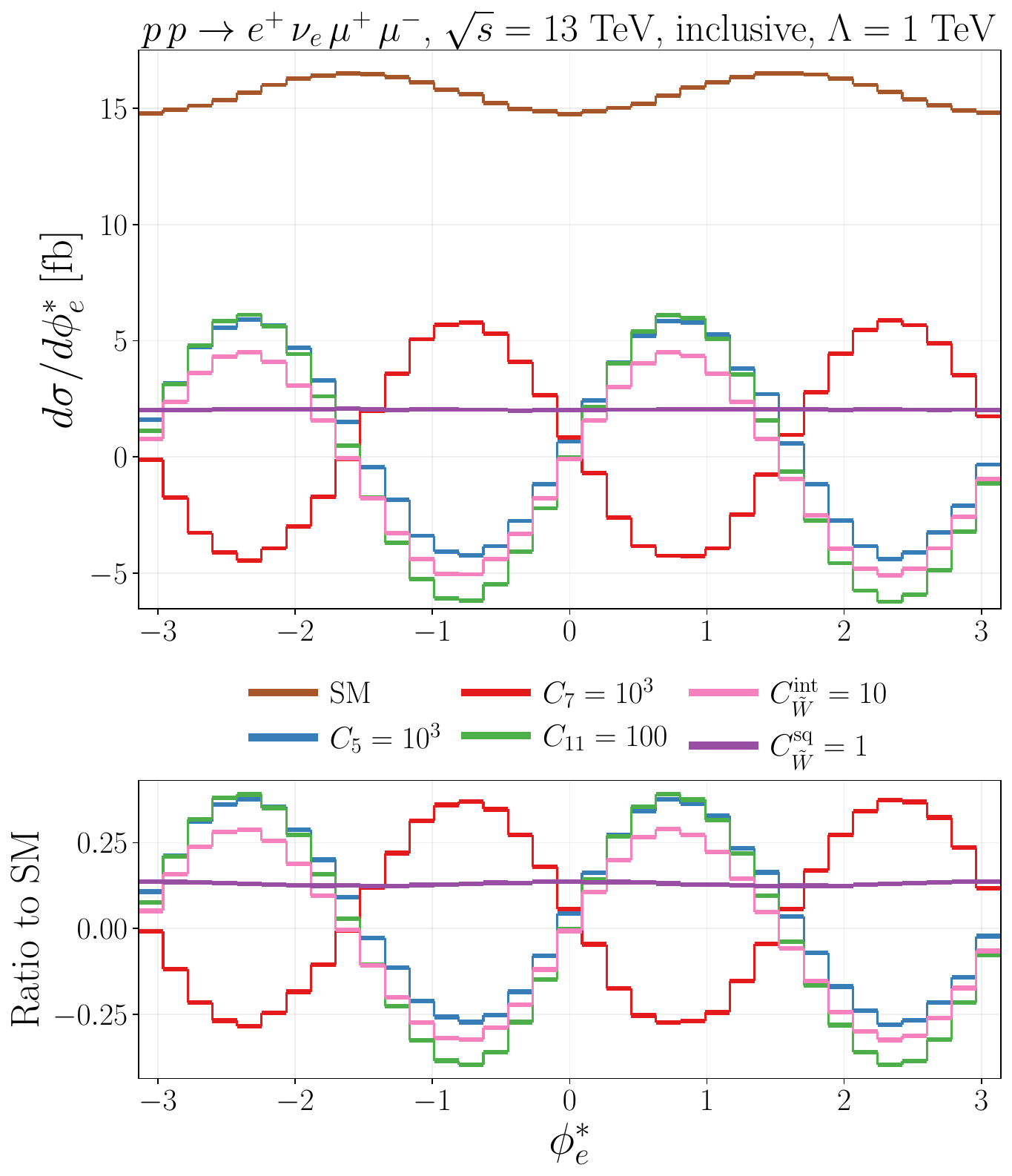}
    \end{subfigure}
    \hfill 
    \begin{subfigure}[b]{0.49\linewidth}
        \centering
       \includegraphics[page=8, width=\textwidth]{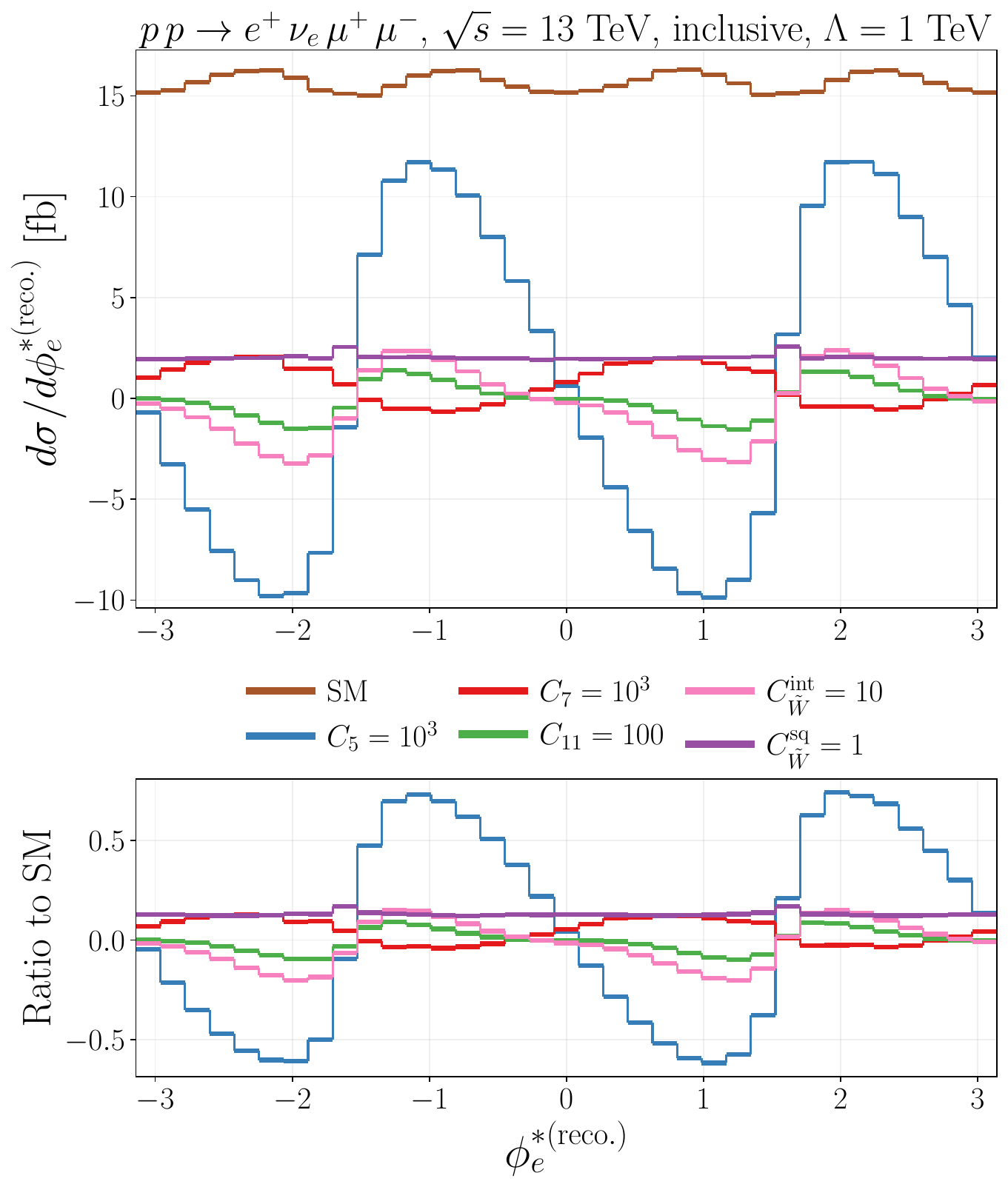}
    \end{subfigure}
    \caption{Differential cross sections in the azimuthal angle of \(e\), including contributions from CP-odd dimension-6 and dimension-8 operators. Both panels correspond to the inclusive setup: the left plot uses truth-level information for the missing transverse momentum, whilst the right plot is based on reconstructed information.}
    \label{fig:wz_cp_odd}
\end{figure}

\begin{figure}[ht]
    \centering
    \begin{subfigure}[b]{0.49\linewidth}
        \centering
         \includegraphics[page=7, width=\textwidth]{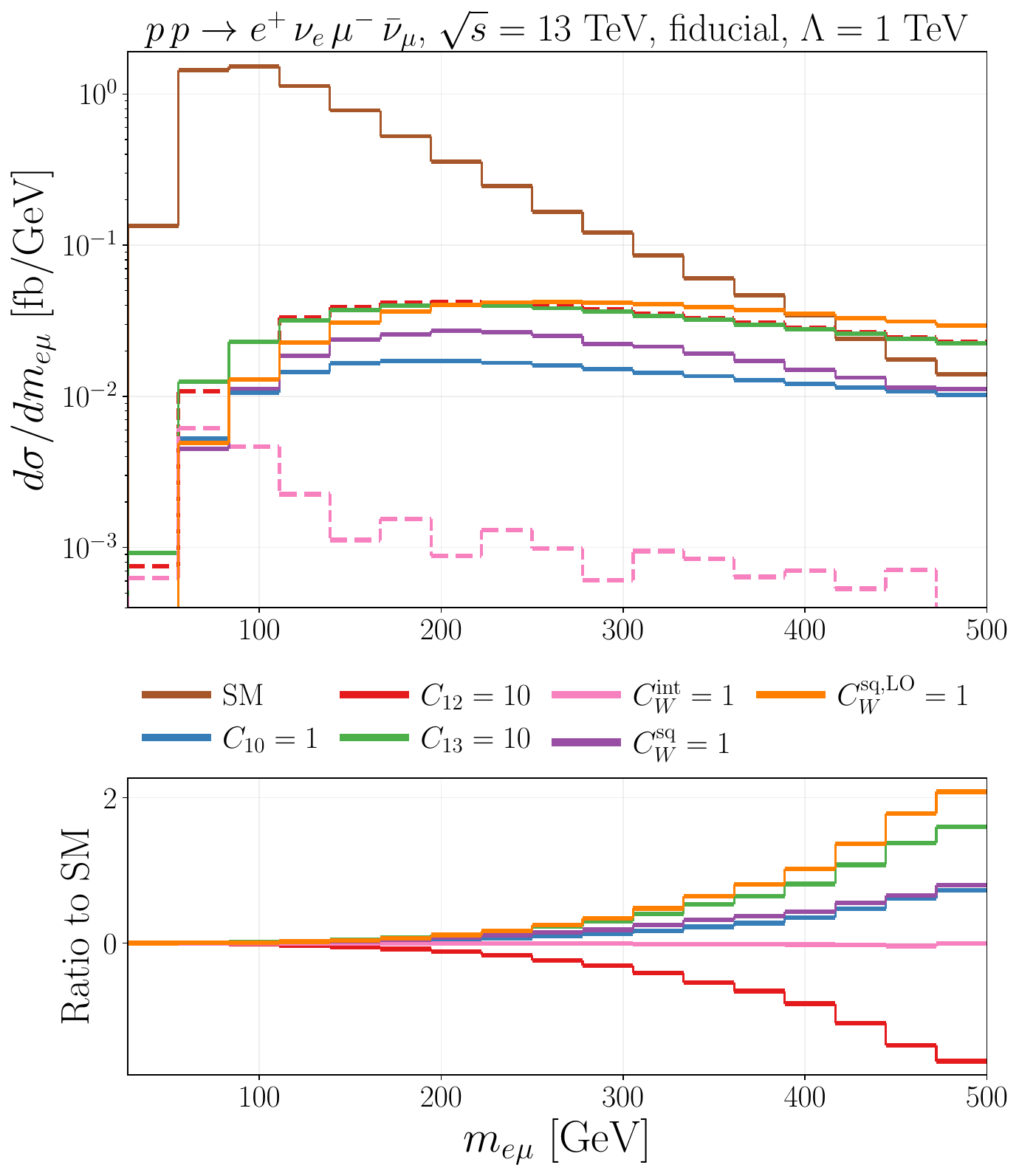}
    \end{subfigure}
    \hfill 
    \begin{subfigure}[b]{0.49\linewidth}
        \centering
        \includegraphics[page=7, width=\textwidth]{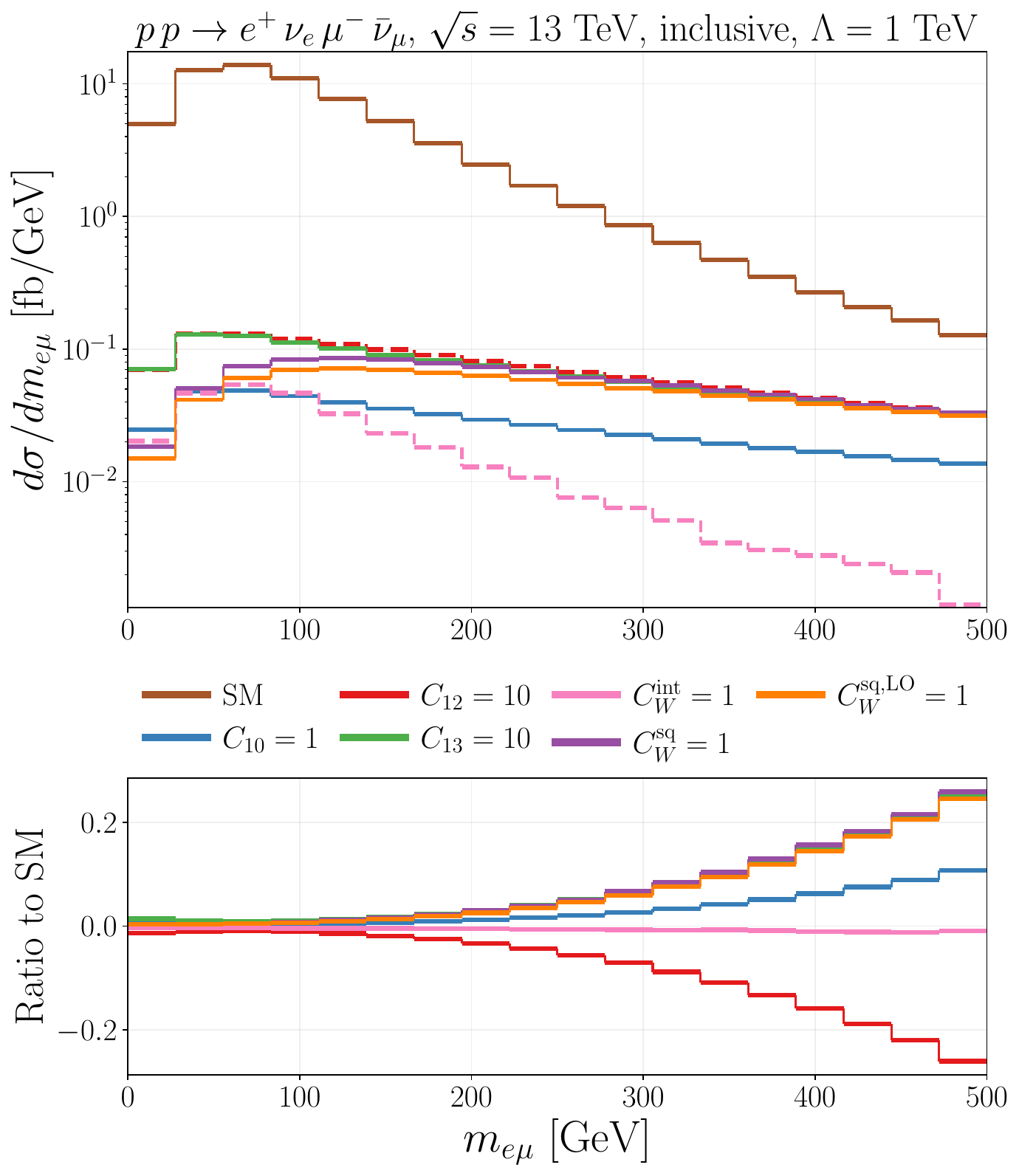}
    \end{subfigure}
    \caption{Distributions of the dilepton invariant mass, $m_{e\mu}$, in $WW$ production in the ATLAS fiducial (left) and the fully inclusive setup (right). The structure of the figure is similar to that of~\cref{fig:wz_fid_m4l_ptz}.}
    \label{fig:ww_fid_fully_incl_mass}
\end{figure}

\begin{figure}[ht]
    \centering
    \begin{subfigure}[b]{0.49\linewidth}
        \centering
         \includegraphics[page=4, width=\textwidth]{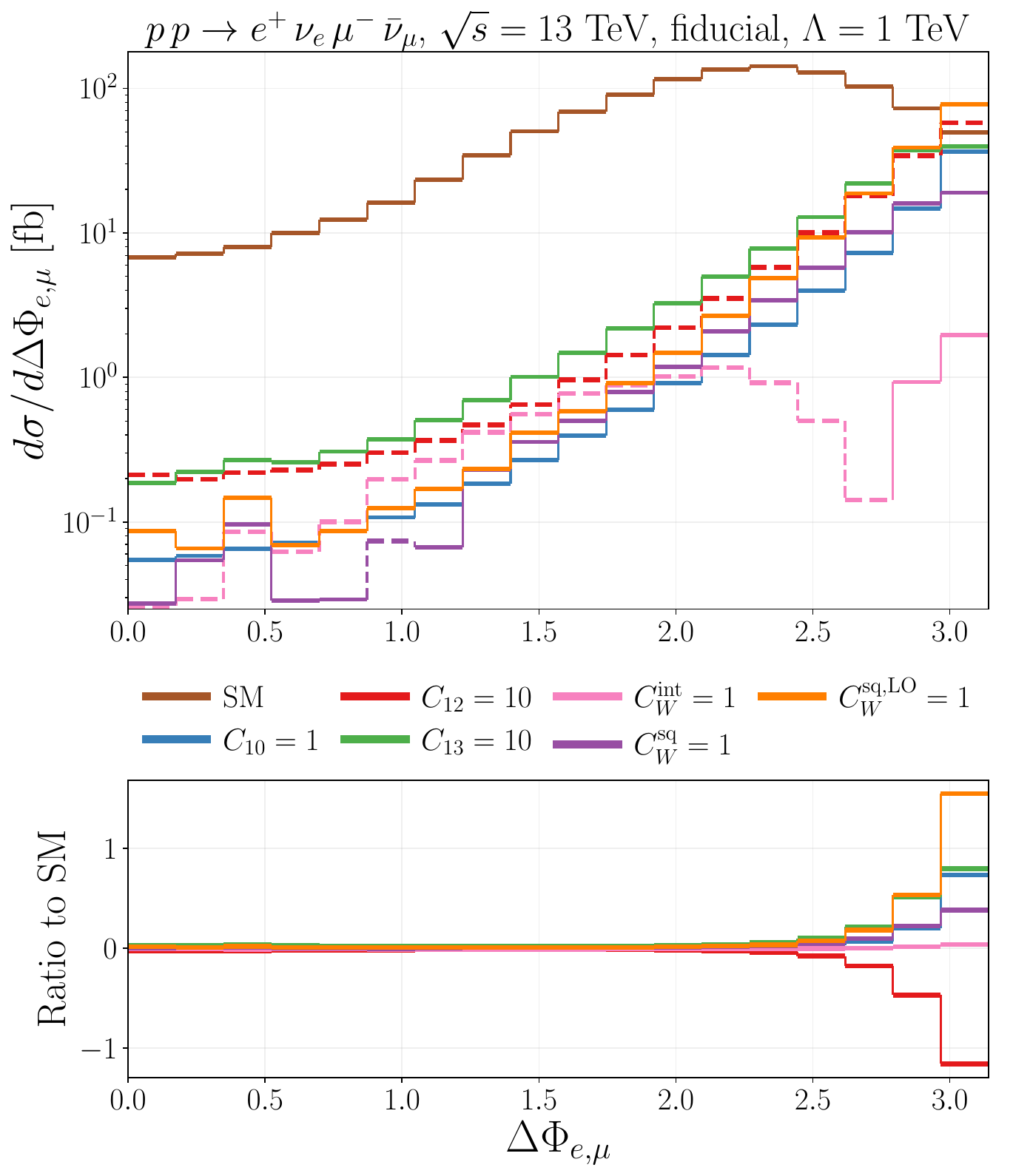}
    \end{subfigure}
    \hfill 
    \begin{subfigure}[b]{0.49\linewidth}
        \centering
        \includegraphics[page=4, width=\textwidth]{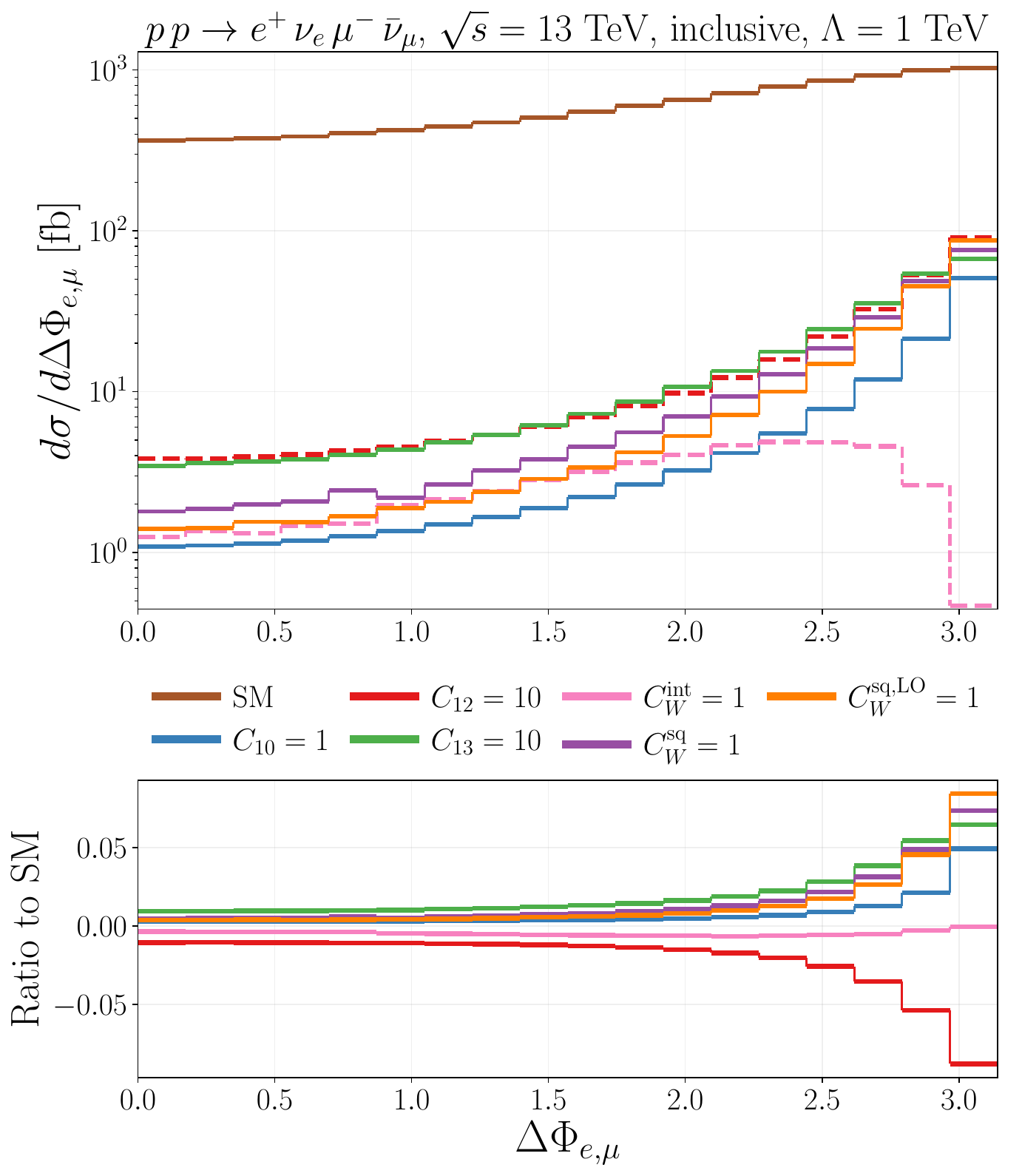}
    \end{subfigure}
    \caption{Same as~\cref{fig:ww_fid_fully_incl_mass} but for the azimuthal separation between the leptons, $\Delta\Phi_{e,\mu}$.}
    \label{fig:ww_fid_fully_incl_dphi}
\end{figure}

In~\cref{fig:wz_fid_m4l_ptz,fig:wz_fid_phi_dphi,fig:wz_fid_cos_dy,fig:wz_cp_odd,fig:ww_fid_fully_incl_mass,fig:ww_fid_fully_incl_dphi}, we present differential predictions for fully leptonic $WZ$ and $WW$ production. The setup of each plot, whether fiducial or inclusive, is indicated at the top of the corresponding panel. In each plot, different colours indicate the SM predictions (SM), interference contributions from the dimension-8 operators denoted as $C_{i}$, with $i \in \{4,5,6,7,10,11,12,13\}$, and the linear and quadratic contributions of the dimension-6 CP-even (CP-odd) operator $\mathcal{O}_{W}$ (with $W \to \tilde{W}$), denoted as $C_{W}^{\text{int}}$ and $C_{W}^{\text{sq}}$, respectively. Dashed lines in the plots indicate negative EFT contributions, i.e. destructive interference. Predictions for the SM and the dimension-6 contributions are computed at NLO in QCD, whilst dimension-8 contributions are evaluated at LO. All EFT contributions are shown for coefficient values chosen to produce deviations comparable to those obtained with the dimension-6 coefficient set to unity.

\paragraph{$\mathbf{WZ}$}
\Cref{fig:wz_fid_m4l_ptz} illustrates distributions in the invariant mass of the four-lepton system (left) and the transverse momentum of the $Z$ boson (right). 
Dimension-8 contributions exhibit a similar energy dependence as dimension-6 squared contributions, though their magnitude is comparatively insignificant—note the up-scaling of the dimension-8 coefficients by at least a factor of 10. It should be stressed that throughout, any comparison between dimension-8 and dimension-6 effects implicitly refers to the squared dimension-6 contribution, as it appears at the same order in the EFT expansion as the linear dimension-8 terms. Nevertheless, for completeness, we also present the linear dimension-6 contribution in all the differential distributions, which is of \(\mathscr{O}(\Lambda^{-2})\).

Moving to~\cref{fig:wz_fid_phi_dphi}, we present the distributions of the azimuthal angle of the anti-muon (left) and the azimuthal separation between the positron and the anti-muon (right). 
As in previous cases, in the angular observables, we also observe mild effects from the dimension-8 operators. Notably, the contributions of the operators $\mathcal{O}_{13}$ and $\mathcal{O}_{12}$—with their corresponding WCs scaled to ten times the value of the dimension-6 coefficient $C_W$—emerges as the most competitive among the dimension-8 contributions across all observables. In the left panel of~\cref{fig:wz_fid_phi_dphi}, we notice that the behaviour of the couples $\mathcal{O}_4, \mathcal{O}_6$ and $\mathcal{O}_{12}, \mathcal{O}_{13}$ differs from the other observables.
Specifically, $\mathcal{O}_4$ and $\mathcal{O}_6$ oscillate with the same frequency, but with different amplitude modulations, whilst the modulation of $\mathcal{O}_{12}$ and $\mathcal{O}_{13}$ differs by an offset. It is worth noting that, whilst the distribution shapes induced by \(\mathcal{O}_{12}\) and \(\mathcal{O}_{13}\) exhibit certain differences, the two coefficients can still display a strong correlation in a joint fit when all bins are considered, owing to the overall symmetry of the distribution across the full range. However, a fit based on a particular combinations of bins breaks the degeneracy, a similar behaviour is observed for \(\mathcal{O}_{4}\) and \(\mathcal{O}_{6}\). In the right panel of \cref{fig:wz_fid_phi_dphi} we observe that the contribution induced by $\mathcal{O}_{4}$ and $\mathcal{O}_{6}$ exhibit a sudden suppression of the cross section as $\Delta\Phi_{e,\mu}$ approaches $\pi$. This behaviour can be attributed to non-trivial cancellations between different regions of phase space where the differential cross-section takes opposite signs—cf. the left panel of~\cref{fig:wz_fid_cos_dy}.

Turning to the polar decay angle distribution in the left panel of~\cref{fig:wz_fid_cos_dy}, we observe a similar pattern to that discussed previously: dimension-8 effects remain generally subdominant, with the operators $\mathcal{O}_{4}$ and $\mathcal{O}_{6}$ exhibiting a sign change in their interference contributions. Finally, in the right panel of~\cref{fig:wz_fid_cos_dy}, the rapidity separation observable exhibits significant shape distortions induced by \(\mathcal{O}_{12}\) and \(\mathcal{O}_{13}\), relative to both the SM and \(\mathcal{O}_W\). This results in distinct two-dimensional shapes in the fit, leading to comparatively enhanced constraining power relative to the other observables—to be discussed later.

Finally, in~\cref{fig:wz_cp_odd}, we show the distributions of the azimuthal angle of the positron, including CP-odd contributions. The left and right panels correspond to predictions obtained using truth-level and reconstructed information on the missing transverse momentum, respectively. Both plots are presented in the inclusive setup. 
CP-odd contributions are generally negligible compared to their CP-even counterparts. Notably, the contribution from \(\mathcal{O}_{11}\) yields an amplitude that is an order of magnitude larger than those from \(\mathcal{O}_5\) and \(\mathcal{O}_7\). However, upon integrating over the full angular domain, cancellations across phase-space regions suppress the total cross section, so the net contributions of all CP-odd operators are of the same order; see~\cref{tab:wz_total_xsec_cp_odd}. 

\paragraph{$\mathbf{WW}$} in contrast to $WZ$ production, the $WW$ channel exhibits a larger cross section but introduces considerable challenges in the fully leptonic decay mode. The dominant top-quark backgrounds produce the same final-state topology as the signal, complicating the isolation of $WW$ events. Furthermore, the two undetected neutrinos greatly restrict the number of measurable differential observables, making it impossible to reconstruct the rest frame of a single $W$ boson. 

In~\cref{fig:ww_fid_fully_incl_mass,fig:ww_fid_fully_incl_dphi}, we present differential predictions for the dilepton invariant mass, $m_{e\mu}$ and the azimuthal separation, $\Delta\Phi_{e,\mu}$, respectively, in the fiducial (left panels) and the fully inclusive (right panels) setups of fully leptonic $WW$ production. The fiducial selection corresponds to the third configuration described earlier for $WW$ production in~\cref{sec:comp_setup}. Unlike in the \( WZ \) analysis, here we also present predictions for dimension-6 effects at LO, for reasons that will be clarified later.

Notably, we observe that the dimension-8 operator $\mathcal{O}_{10}$ exhibits a competitive effect compared to the quadratic dimension-6 contribution, particularly pronounced in the back-to-back regions and in the high-energy tails. However, this effect is more evident in the fiducial setup. The reason can be attributed to the jet-veto effect in the fiducial setting. As has also been concluded in previous studies, see for e.g.~\cite{ElFaham:2024uop}, and whilst the jet-veto application is necessary for the background suppression in $WW$ production, the quadratic EFT contributions at NLO are significantly smaller than the LO ones. This is due to the jet veto significantly suppressing the NLO real emissions.
Such an effect is crucial in this study, whose primary objective is to compare the quadratic contributions from dimension-6 to the linear contributions from dimension-8 operators. 
Moreover, it also motivates the study of dimension-8 effects at NLO in QCD which we postpone for future work.

Finally, we observe that the contributions from \( \mathcal{O}_{12} \) and \( \mathcal{O}_{13} \) exhibit similar behaviour. This is expected, as the leading high-energy contribution of those operators to the interference with the SM scales with the linear combination \( (C_{12} - C_{13}) \)~\cite{Degrande:2023iob}. As a result, any differences between the two operators arise from subleading terms, which become relevant only in the low-energy bins of the distributions. This observation holds for both \(WW\) and \(WZ\), although it is more pronounced in the latter.

\subsection{Polarisation fractions}
\label{sec:pol_frac}
We now shift focus to angular observables in boson decays, which are particularly sensitive to interference effects and probe the polarisation of the gauge bosons. In the absence of any selection cuts on the leptons, and assuming access to MC truth-level momenta, i.e. without neutrino reconstruction, the differential decay rate of an electroweak boson into two final-state particles can be expressed as follows:
\beqn
\frac1{\sigma}\frac{\rd\sigma}{\rd\cos\theta^* \,\rd\phi^*} 
&=&
\frac{3}{16\pi}
\bigg[
1+\cos^2\theta^*  + \frac{A_0}{2}(1-3\cos^2\theta^*) + A_1 \sin2\theta^* \cos\phi^* \nonumber\\
&&\hspace*{0.8cm}+\frac12 A_2 \sin^2\theta^* \cos2\phi^*  + A_3 \sin\theta^* \cos\phi^* 
+ A_4\cos\theta^*  \nonumber\\
&&\hspace*{0.8cm}+ A_5 \sin\theta^* \sin\phi^* 
+ \; A_6  \sin2\theta^* \sin\phi\;+  A_7 \sin^2\theta^* \sin2\phi^* \bigg]\,,
\eeqn
where $\phi^*$ and $\theta^*$ are, respectively, the azimuthal and polar angles of one of the decay leptons. Upon integration over the azimuthal decay angle, the decay rate reads: 
\begin{equation}
\frac{1}{\sigma}\,\frac{\mathrm{d}\sigma}{\mathrm{d}\cos\theta^{*}}
= \frac{3}{8}\left( 1 + \cos^{2}\theta^{*}
+ \frac{A_{0}}{2}\left(1 - 3\cos^{2}\theta^{*}\right)
+ A_{4}\cos\theta^{*} \right).
\label{eq:costhetaA0A4}
\end{equation}
Through linear combinations of the coefficients $A_0,\,A_4$, this expression can be re-written in terms of polarisation fractions:
\beqn
\frac{1}{\sigma} \frac{\rd \sigma}{\rd\, \textrm{cos}\theta^*} 
&=&
\frac38\bigg[
\,\,2\,f_{\rm 0}\,\sin^2\theta^*\nonumber\\
&&\hspace*{0.5cm}+f_{\rm L}\,\left(1+\cos^2\theta^{*}-2\,c_{\rm \tiny LR}\,\cos\theta^*\right) \,\nonumber\\
&&\hspace*{0.5cm}+f_{\rm R}\,\left(1+\cos^2\theta^{*}+2\,c_{\rm LR}\,\cos\theta^*\right) \, \bigg], \label{eq:polfrac}
\eeqn
where $f_{\rm 0}, f_{\rm R}, f_{\rm L}$ are the longitudinal, right-handed, and left-handed polarisation fractions, respectively. The parameter \( c_{\rm LR} \) characterises the interplay between left- and right-handed couplings of the EW boson to massless leptons, effectively quantifying their relative contribution;
\beq\label{eq:CLR}
c_{\rm LR} = \frac{|g_{\rm L}|^2-|g_{\rm R}|^2}{|g_{\rm L}|^2+|g_{\rm R}|^2}\,,
\eeq
which are unaffected by the SMEFT operators considered in this work and are therefore fixed to their SM values:
\beq
c_{\rm LR}^{W} = 1\,,\qquad c_{\rm LR}^{Z} = \frac{1-4\sin^2\theta_{W}}{1-4\sin^2\theta_{W}+8\sin^4\theta_{W}}\,\approx 0.215\,,\quad \sin^2\theta_{W} = 1-\frac{m_{W}^2}{m_{Z}^2}\,.
\eeq
The $A_0$ and $A_4$ coefficients are extracted from~\cref{eq:costhetaA0A4} and they are directly related to the 
longitudinal- and transverse-polarisation fractions through the following relations:
\beq\label{eq:Atof_corresp}
A_0 = 2\,f_0\,,\qquad A_4 = 2\,c_{\rm LR}\,(f_{\rm R}-f_{\rm L})\,.
\eeq

\begin{figure}[ht]
    \centering
    \begin{subfigure}{0.5\textwidth}
        \centering
        \includegraphics[width=\linewidth]{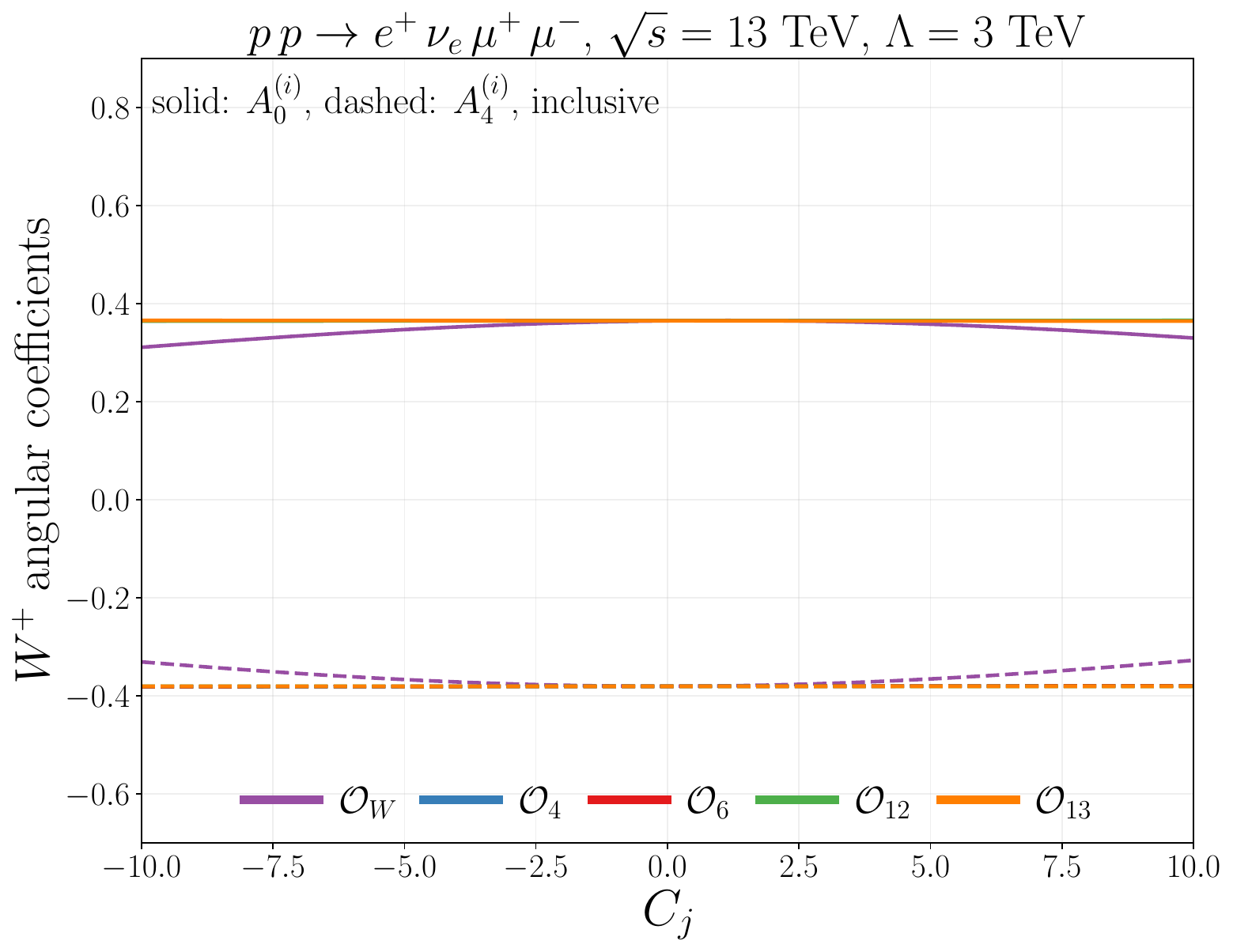}
    \end{subfigure}%
    \begin{subfigure}{0.5\textwidth}
        \centering
        \includegraphics[width=\linewidth]{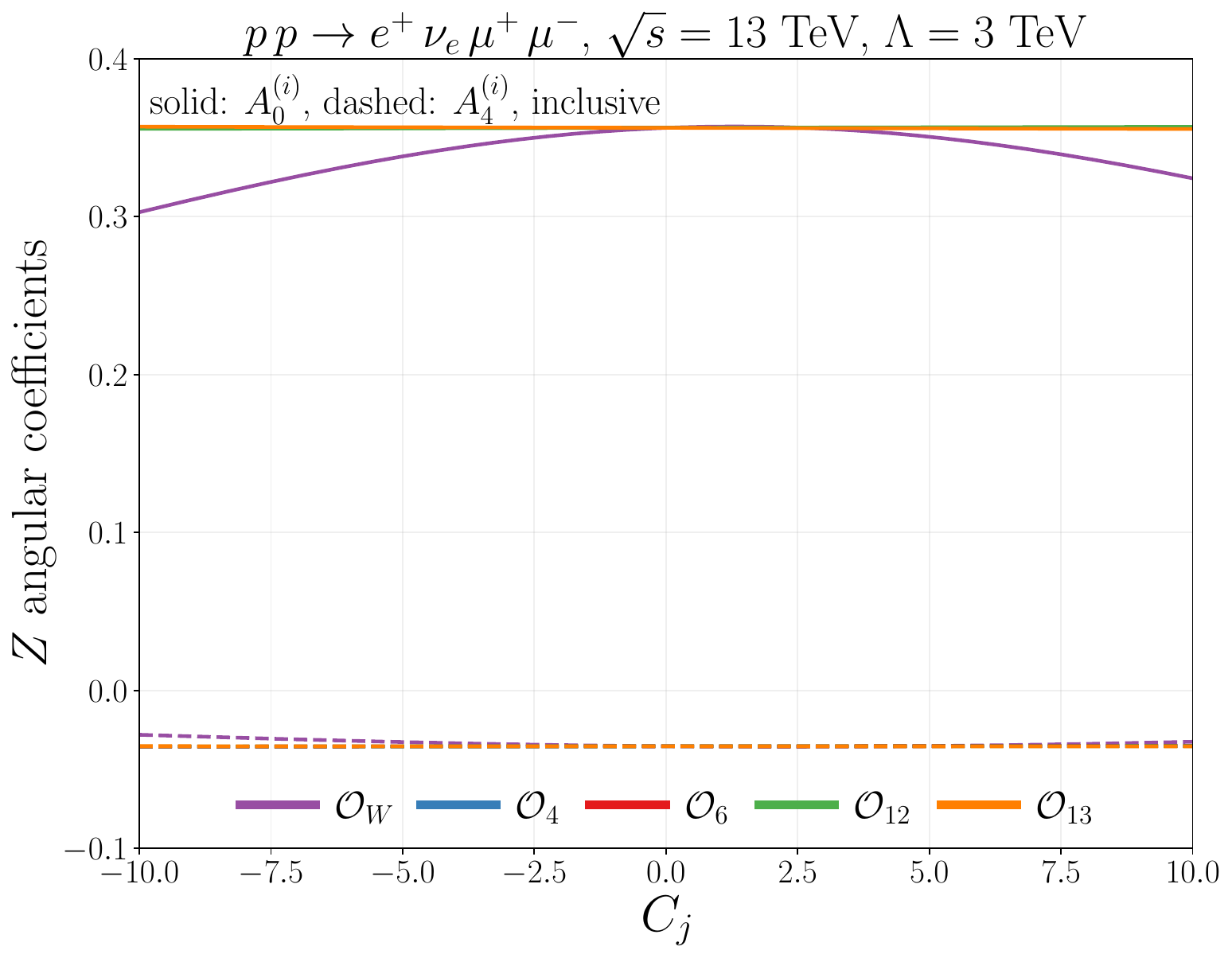}
    \end{subfigure}

    \begin{subfigure}{0.5\textwidth}
        \centering
        \includegraphics[width=\linewidth]{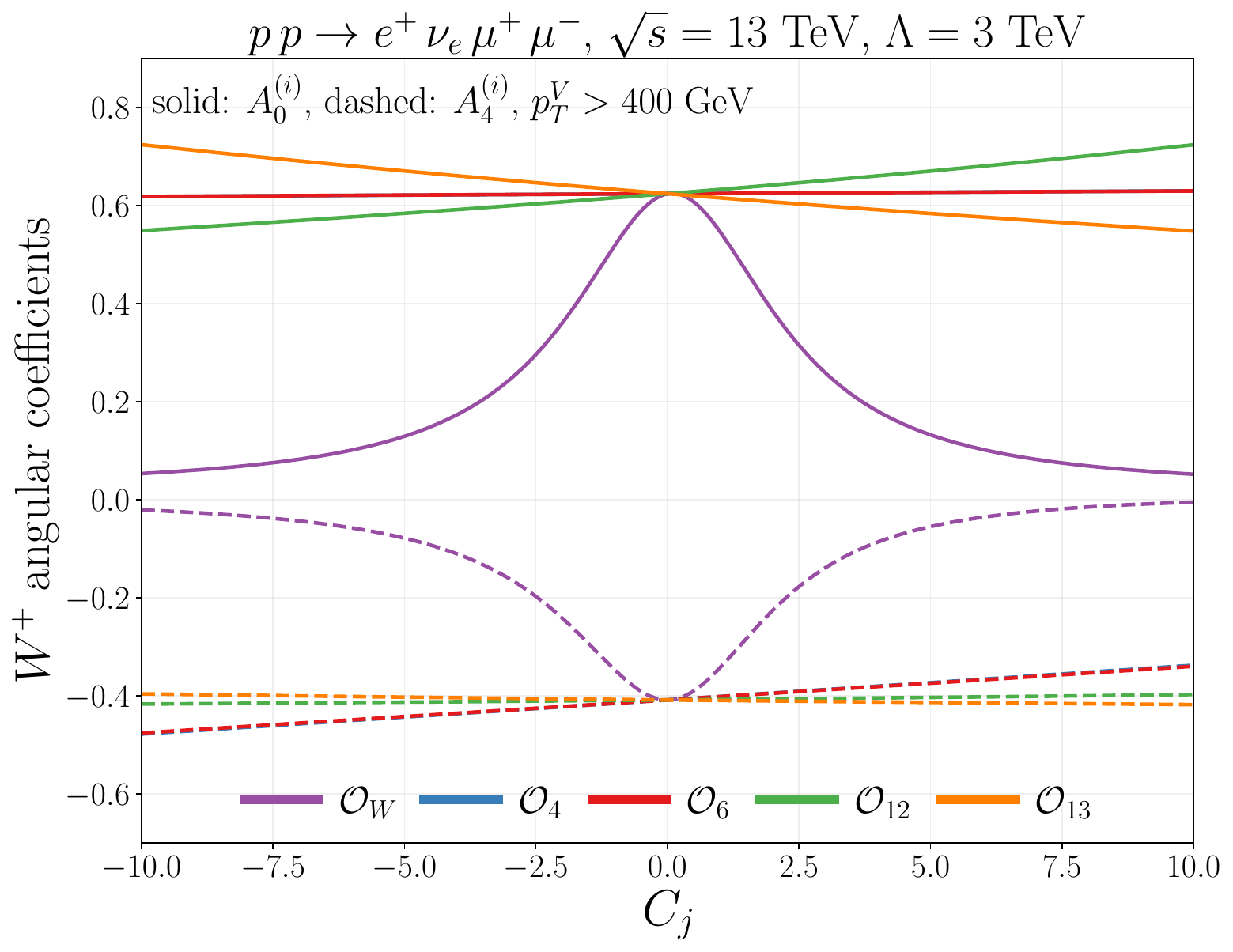}
    \end{subfigure}%
    \begin{subfigure}{0.5\textwidth}
        \centering
        \includegraphics[width=\linewidth]{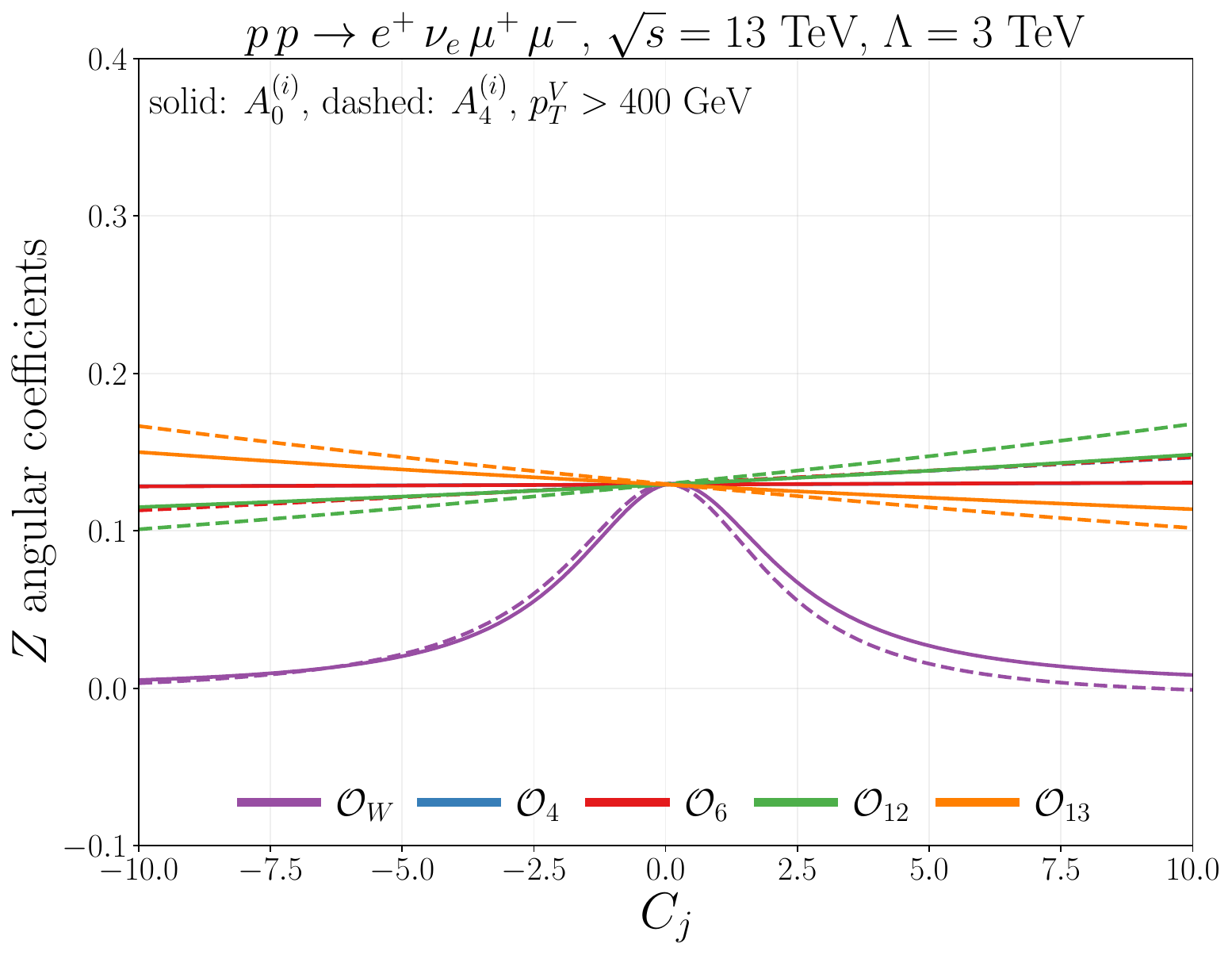}
    \end{subfigure}
    \caption{Dependence of the angular coefficients \(A_{0}\) (solid) and \(A_{4}\) (dashed) for the \(W\) boson (left) and the \(Z\) boson (right) in \(WZ\) production on the Wilson coefficients of CP-even operators. Top panels are computed for an inclusive setup, while bottom panels are computed with a minimal cut on the $p_T$ of the EW boson of $400$ GeV. The parametrisation is given in terms of \(A_i^{(2)}\), shown for the dimension-6 operator \(\mathcal{O}_W\) including both linear and quadratic terms, and \(A_i^{(1)}\), shown for the dimension-8 operators retaining only the linear contributions. }
    \label{fig:angular_coeff}
\end{figure}

The angular coefficients \(A_{0}\) and \(A_{4}\), and consequently, the corresponding polarisation fractions, are shown in~\cref{fig:angular_coeff} as functions of all CP-even dimension-6 and dimension-8 coefficients relevant to \(WZ\) production. The corresponding CP-odd results are negligible due to the numerically insignificant contributions of the considered SMEFT operators and therefore, are not shown. The dependence of the angular coefficients is parametrised as follows. For a given angular coefficient \(A_i\), two truncations of the SMEFT expansion are considered: one at linear order and one including quadratic terms—with the dimension-8 contribution entering at \(\mathscr{O}(\Lambda^{-4})\):  
\begin{equation}\label{eq:coef_par}
\begin{aligned}
A^{(1)}_i(\lambda_j) &= 
  \frac{ A_i^{\rm SM} + \lambda_j \, A_i^{\rm int}\,\kappa^{\rm int} }
       { 1 + \lambda_j \,\kappa^{\rm int} } \,, 
&
A^{(2)}_i(\lambda_j) &= 
  \frac{ A_i^{\rm SM} + \lambda_j \, A_i^{\rm int}\,\kappa^{\rm int} + \lambda_j^2 \, A_i^{\rm sq}\,\kappa^{\rm sq} }
       { 1 + \lambda_j \,\kappa^{\rm int} + \lambda_j^2 \,\kappa^{\rm sq} }\,,
\end{aligned}
\end{equation}
where
\beq\label{eq:kappadef}
  \kappa^{\rm int} = \frac{\sigma^{\rm int}}{\sigma^{\rm SM}}\,, 
  \qquad 
  \kappa^{\rm sq}  = \frac{\sigma^{\rm sq}}{\sigma^{\rm SM}}\,, 
  \qquad 
  j \in \{W,4,6,12,13\}\, .
\eeq
and we defined
\begin{equation}
    \lambda_W = \frac{C_W}{\Lambda^2}, \qquad \lambda_n = \frac{C_n}{\Lambda^4}, \ n \in \{4,6,12,13\}.
\end{equation}
The results are reported for \( WZ \) production in an inclusive setup, using truth-level MC information for the missing transverse momentum. We emphasise that extracting these coefficients via spherical–harmonics projections, i.e.~\cref{eq:costhetaA0A4}, is well defined only without fiducial cuts on decay products and with perfect neutrino reconstruction, which allows full integration over the decay angles. In a fiducial setup with realistic neutrino reconstruction, applying the same strategy can yield unphysical results~\cite{Boudjema:2009fz,Stirling:2012zt,Ballestrero:2017bxn,Denner:2020bcz}.

We consider two different setups. The first is a fully inclusive configuration and the second applies a lower cut on the transverse momentum of the EW boson, $p_T^V > 400$ GeV. In the second case, we focus on the boosted regime in which the effects of SMEFT operators are more pronounced. The results are shown in \cref{fig:angular_coeff} for a new physics scale of $\Lambda = 3$ TeV, at which quadratic contributions of the dimension-8 operators are negligible.

In the top panel of \cref{fig:angular_coeff}, which shows results for the inclusive setup, we observe that the contributions from dimension-8 operators to the angular coefficients are negligible, whilst the dimension-6 operators induce only small deviations. This behaviour is a consequence of the suppressed contributions of $\kappa^{\rm int}$ and $\kappa^{\rm sq}$ when $\Lambda = 3$ TeV. As a result, the polarisation fractions remain very close to the SM prediction.

In the bottom panel, for the boosted scenario, we observe that the dimension-6 operator has a large impact and the contributions from dimension-8 operators become non-negligible. In both the left and right panels, and focusing on $A_0^{(1)}$ for the dimension-8 operators, the slope of the line is dominated by the denominator of~\cref{eq:coef_par}. This indicates that the angular coefficient $A_0^{\mathrm{int}}$ which corresponds to the interference of the dimension-8 amplitude with the SM is negligible. This observation is consistent with the findings of Ref.~\cite{Degrande:2023iob}, where the corresponding helicity amplitudes were studied and it was shown that the production of longitudinal vector bosons is suppressed 
compared to transverse ones, vanishing in the case of the operators $\mathcal{O}_{12}$ and $\mathcal{O}_{13}$ for the production of two longitudinal bosons. Turning to $A_4^{(1)}$, we find a different behaviour: as the vector bosons produced with an insertion of the dimension-8 operators are mostly transversely polarised, their contribution to the angular coefficient $A_4$ is mild, but not negligible. We also comment on the dimension-6 operator  which leads mostly to transverse bosons, with $A_0$ and thus $f_0$ tending to zero for large coefficients where the EFT contribution dominates. 

Furthermore, we note the degeneracy between the coefficients $(C_4, C_6)$ and $(C_{12}, C_{13})$, consistent with what was already observed in the differential distributions. This degeneracy is most apparent in the bottom panel. Finally, concerning the $Z$-boson panel (right), we observe that, unlike in the $W$ case, the $Z$ boson exhibits nearly equal left- and right-handed polarisation fractions in the inclusive setup, which results in a much smaller value of $A_{4}$, as expected. Moving to the boosted setup, and focusing on the right panel, we observe that the values of $A_0$ and $A_4$ become very close, with $A_0$ being suppressed and $A_4$ enhanced. This leads to a reduction in the longitudinal polarisation fraction of the $Z$ boson, and to an excess of right-polarised over left-polarised bosons.

\section{Sensitivity study}
\label{sec:fit}
Using a statistical approach based on a $\chi^2$ analysis, we combine experimental measurements and SMEFT predictions to derive constraints on the Wilson coefficients considered in this work. We employ differential data from multiple observables and kinematic regions. Angular observables that are sensitive to the polarisation structure of the diboson events are also included. The total experimental uncertainties are computed as the quadrature sum of statistical errors and systematic uncertainties. Since we assume uncorrelated errors, we introduce an additional 5\% systematic uncertainty, applied bin-by-bin, as a conservative measure.

We adopt SM predictions as used in ATLAS analyses: NNLO QCD for the $WZ$~\cite{Grazzini:2016ctr} and NNLO QCD combined with NLO EW corrections for $WW$ production~\cite{Grazzini:2017ckn,Grazzini:2019jkl}. Within the fiducial $WZ$ region under consideration~\cite{ATLAS:2019bsc,ATLAS:2022oge}, NLO EW corrections amount to about $-3\%$ relative to NNLO QCD results~\cite{Grazzini:2019jkl,Le:2022lrp}; thus, their omission is unlikely to significantly affect our fits. The SM predictions are extracted by digitising plots presented in experimental papers, as numerical results were not available in the HEPData database~\cite{Maguire:2017ypu}. 

By performing a $\chi^2$ fit, we establish individual and marginalised 95\% confidence level (CL) intervals for the WCs, quantifying the agreement of the SMEFT predictions with data. Our bounds apply exclusively within experimentally defined fiducial regions. Our EFT theoretical predictions are computed as described in~\cref{sec:comp_setup}.

Our analysis considers the EFT parametrisation up to $\mathcal{O}(\Lambda^{-4})$. It is worth noting that we have explicitly checked the contributions arising from the squared terms of dimension-8 operators. We find that, for coefficients of order one to ten, i.e.\ $C = \mathcal{O}(1\text{--}10)$, and in the region where the cut-off scale is of order $3~\text{TeV}$ or higher such contributions are subdominant.

\paragraph{$\mathbf{WZ}$}
we make use of the fiducial differential cross section data provided by the ATLAS collaboration~\cite{ATLAS:2019bsc, ATLAS:2022oge}. The considered differential measurements include the following observables:
\beq
\{p_{T}^Z,\quad
p_{T}^W,\quad
m_{T}^{WZ},\quad
\Delta\Phi_{W,Z},\quad
p_{T}^{\nu},\quad
|\Delta y_{Z,l_{W}}|,\quad
\cos\theta^*_{e}\,\}.
\eeq
For the polar decay angle, $\cos\theta^*_{e}$, we use EFT predictions corresponding to its reconstructed definition, following the neutrino reconstruction method implemented by ATLAS~\cite{ATLAS:2019bsc}. We highlight that all differential observables, with the exception of $\cos\theta^*_{e}$, are taken from Ref.~\cite{ATLAS:2019bsc}, based on data corresponding to an integrated luminosity of 36.1~fb$^{-1}$. In contrast, the polar decay angle data are extracted from Ref.~\cite{ATLAS:2022oge}, which uses a larger dataset with an integrated luminosity of 139~fb$^{-1}$. 

\paragraph{$\mathbf{WW}$}
we employ the differential measurements reported by ATLAS in Ref.~\cite{ATLAS:2019rob}. The analysis incorporates several kinematic distributions:
\beq
\{p_{T}^{l_{\rm lead.}},\quad
p_{T}^{e\mu},\quad
m_{e\mu},\quad
\Delta\Phi_{e,\mu},\quad
|y_{e\mu}|,\quad
\left|\tanh\frac{\Delta\eta_{e,\mu}}{2}\right|\,\},
\eeq
from which we extract bounds using the first three measurements. The remaining distributions exhibit systematic discrepancies between data and SM predictions—specifically, a consistent underestimation of the fiducial cross sections by 15–20\% as noted in Ref.~\cite{ATLAS:2019rob}. Due to this modelling tension, we do not consider the resulting bounds from these observables reliable for constraining SMEFT effects and therefore exclude them from our results. 

\paragraph{HL-LHC projections} experimental central values are assumed to coincide with the SM predictions. Statistical uncertainties are rescaled according to the luminosity, whilst systematic uncertainties are uniformly reduced by a factor of two. 

\subsection{Fit method and results}
Throughout our analysis, all operator coefficients are scanned within the range \( C_i \in [-10,\,10] \) to remain within the perturbative regime but allowing also strongly coupled UV complete models\footnote{More sophisticated priors on the Wilson coefficients can be applied by employing arguments based on particular UV dynamics, see for example Refs. \cite{Contino:2016jqw,Liu:2016idz}. }. At cut-off scales \(\Lambda \geq 3~\text{TeV}\), the resulting dimension-8 constraints become weaker than the imposed prior and are therefore regarded as unconstrained, with the exception of the coefficient \(C_{10}\) in the \(WW\) channel. 
The corresponding bounds on \(C_{10}\) in \(WW\) production at \(\Lambda = 3~\text{TeV}\), 
extracted from the three observables under consideration and based on current LHC data, 
are reported in~\cref{tab:ww_all_bounds}. Results for different values of \(\Lambda\) 
are presented in~\cref{tab:ww_cw_lambda_bounds_marg,tab:ww_cw_lambda_bounds_ind} 
of~\cref{app:add_results}.
\begin{table}[ht]
    \renewcommand{\arraystretch}{1.25}
    \centering
    \begin{tabular}{l|c|c|c|c}
          \hline obs & $C_W^{\rm LO}$, ind  & $C_W$, ind & $C_{10}$, ind &  $C_W$, marg 
          \\  \hline 
          $p_T^{{l}_{\rm lead.}}$ 
          & [-1.45, 1.42] 
          & [-2.60, 2.81] 
          & [-2.50, 7.41] 
          & [-4.33, 4.54] 
          \\ 
          $p_T^{e\mu}$ 
          & [-1.63, 1.58] 
          & [-3.13, 3.30] 
          & -
          & [-4.09, 4.23] 
          \\ 
          $m_{e\mu}$ 
          & [-1.96, 1.96] 
          & [-4.73, 5.00] 
          & -
          & [-6.78, 7.04] 
          \\ \hline 
    \end{tabular}
    \caption{Individual (ind) bounds on the dimension-6 coefficient \(C_{W}\) and the dimension-8 coefficient \(C_{10}\) in the \(WW\) channel, shown for each of the observables (obs) considered with current LHC data. For the dimension-6 coefficient, bounds are presented at LO and NLO. The last column reports the marginalised (marg) bounds for $C_{W}$ at NLO. All bounds are extracted at $\mathscr{O}(\Lambda^{-4})$ and assume a cut-off scale of $\Lambda = 3~\text{TeV}$. The `–' indicates the coefficient is unconstrained within the imposed prior.}
    \label{tab:ww_all_bounds}
\end{table}

In~\cref{tab:ww_all_bounds}, the looser bounds at NLO can be attributed to jet-veto effects in the \(WW\) channel, as discussed in Ref.~\cite{ElFaham:2024uop}, and can be understood as a consequence of the less-than-unity \(K\)-factors in the fiducial setup shown in~\cref{fig:ww_fid_fully_incl_mass}. As previously mentioned, the coefficient \(C_{10}\) receives a meaningful individual constraint from $p_{T}^{l_{\rm lead.}}$ even for $\Lambda=3$~TeV. This observation is consistent with the findings of Ref.~\cite{Degrande:2023iob}, which emphasise the importance of the operator $\mathcal{O}_{10}$. Finally, we note that the marginalised bounds on \(C_W\) at NLO differ significantly from the individual ones; this feature also appears in the \(WZ\) channel—marginalised and individual bounds from the latter are presented in~\cref{tab:wz_cw_lambda_bounds_marg,tab:wz_cw_lambda_bounds_ind} of~\cref{app:add_results}—and will be discussed in detail in the following section. 

Our results agree with the findings of Ref.~\cite{Mantani:2025bqu}. We find that, even in a quadratic analysis of diboson observables, it is not possible to simultaneously remain within the weakly coupled regime and satisfy the EFT validity condition. When ensuring that the energy probed is significantly smaller than the cut-off scale $\Lambda$, the bounds obtained on $C_W$ enter in the strong coupling regime.  

\subsection{Impact of dimension-8 operators in the fit}
\begin{figure}[ht]
    \centering
    \includegraphics[width=0.7\linewidth]{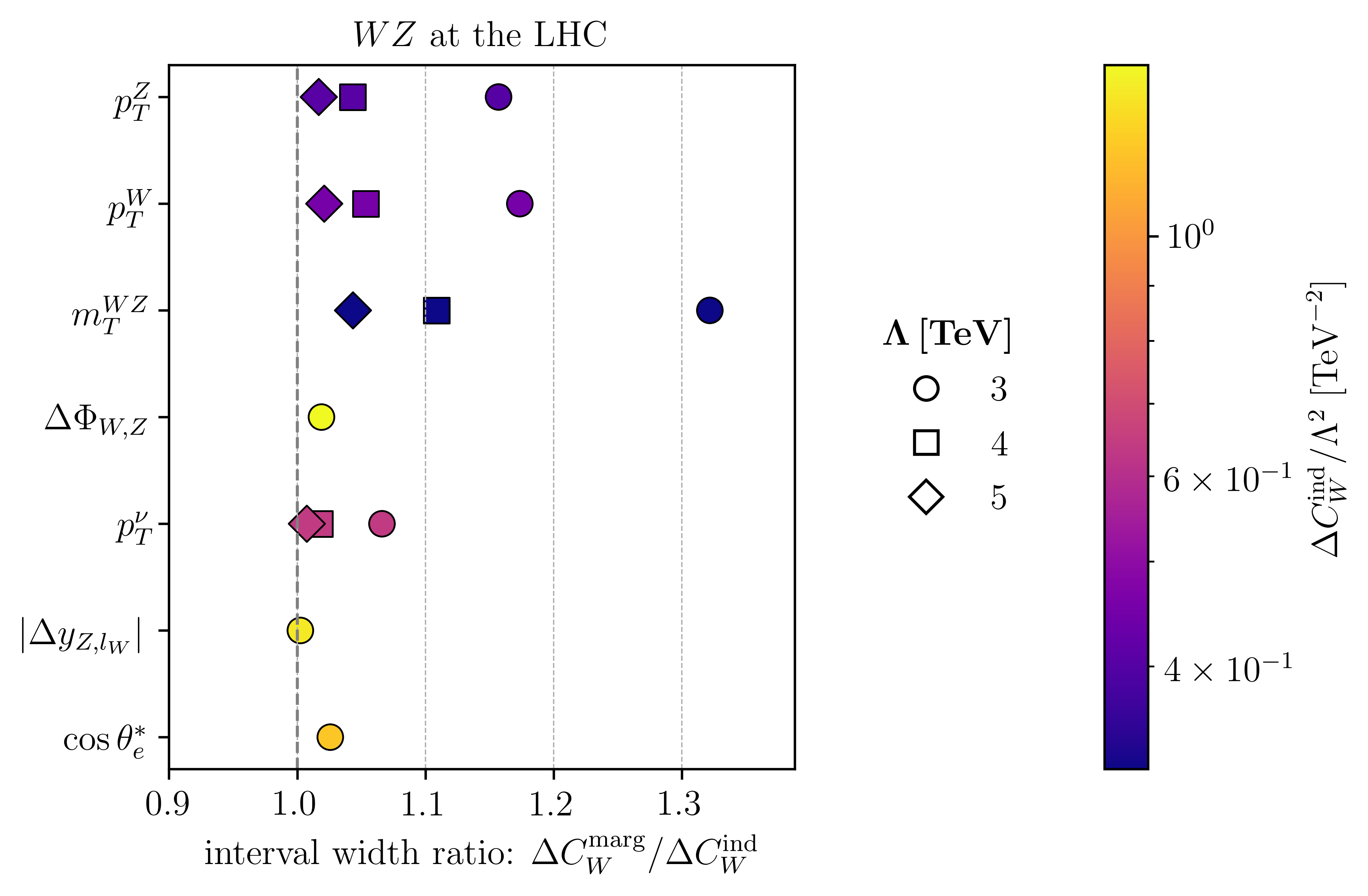}
    \caption{Shown for each observable at different new-physics scales $\Lambda$—denoted by the different marker shapes—is the ratio between the widths of the marginalised and individual bounds of the dimension-6 coefficient $C_W$ at NLO. The colour scale indicates the stringency of the individual bounds. All results are at $\mathscr{O}(\Lambda^{-4})$ in the EFT parametrisation.}
    \label{fig:wz_bounds_ratio}
\end{figure}

\begin{figure}[ht]
    \centering
    \includegraphics[width=0.7\linewidth]{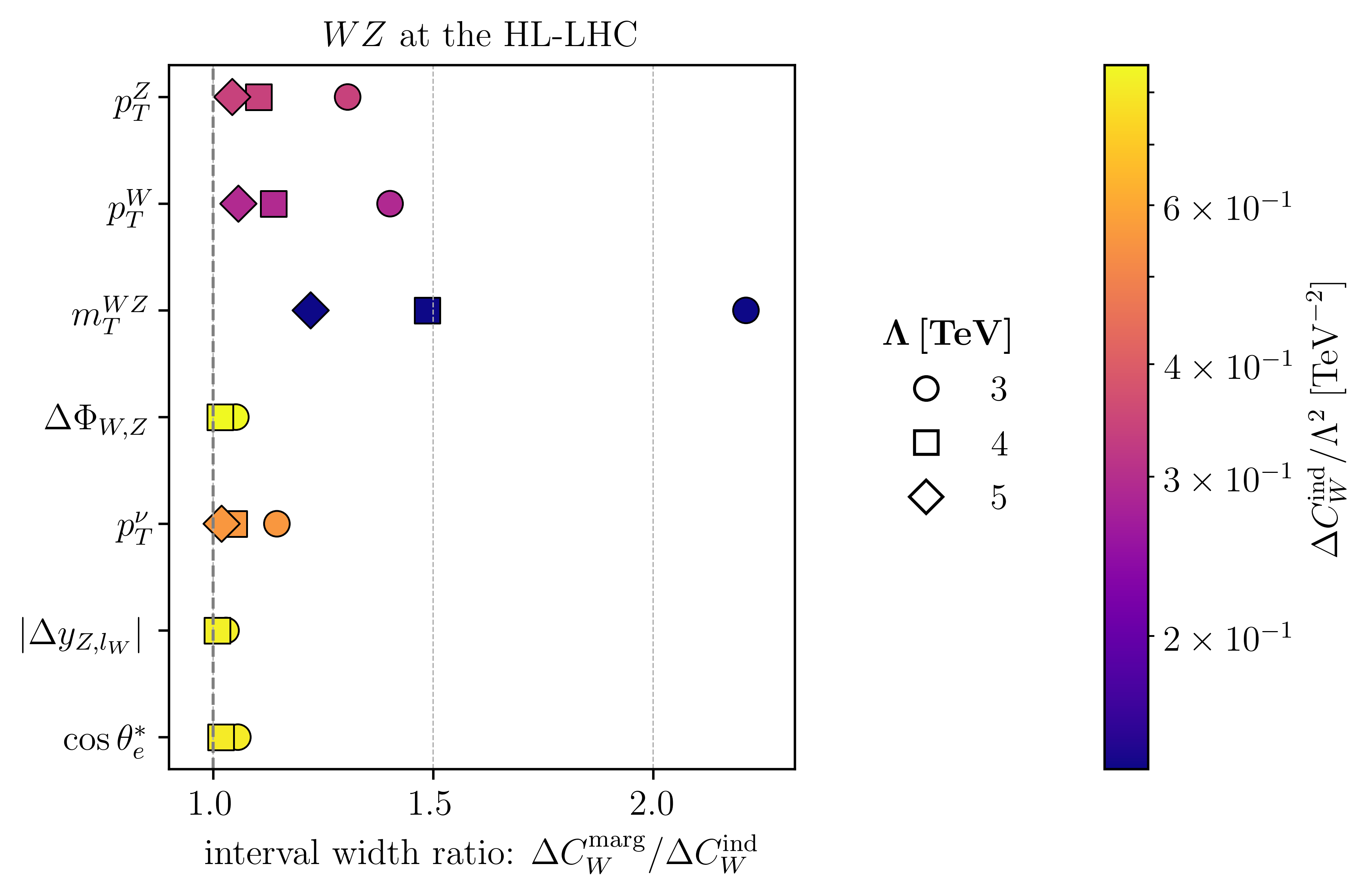}
    \caption{Same as~\cref{fig:wz_bounds_ratio} but for the HL-LHC projections.}
    \label{fig:wz_bounds_ratio_HL}
\end{figure}

\begin{figure}[ht]
    \centering
    \includegraphics[width=0.7\linewidth]{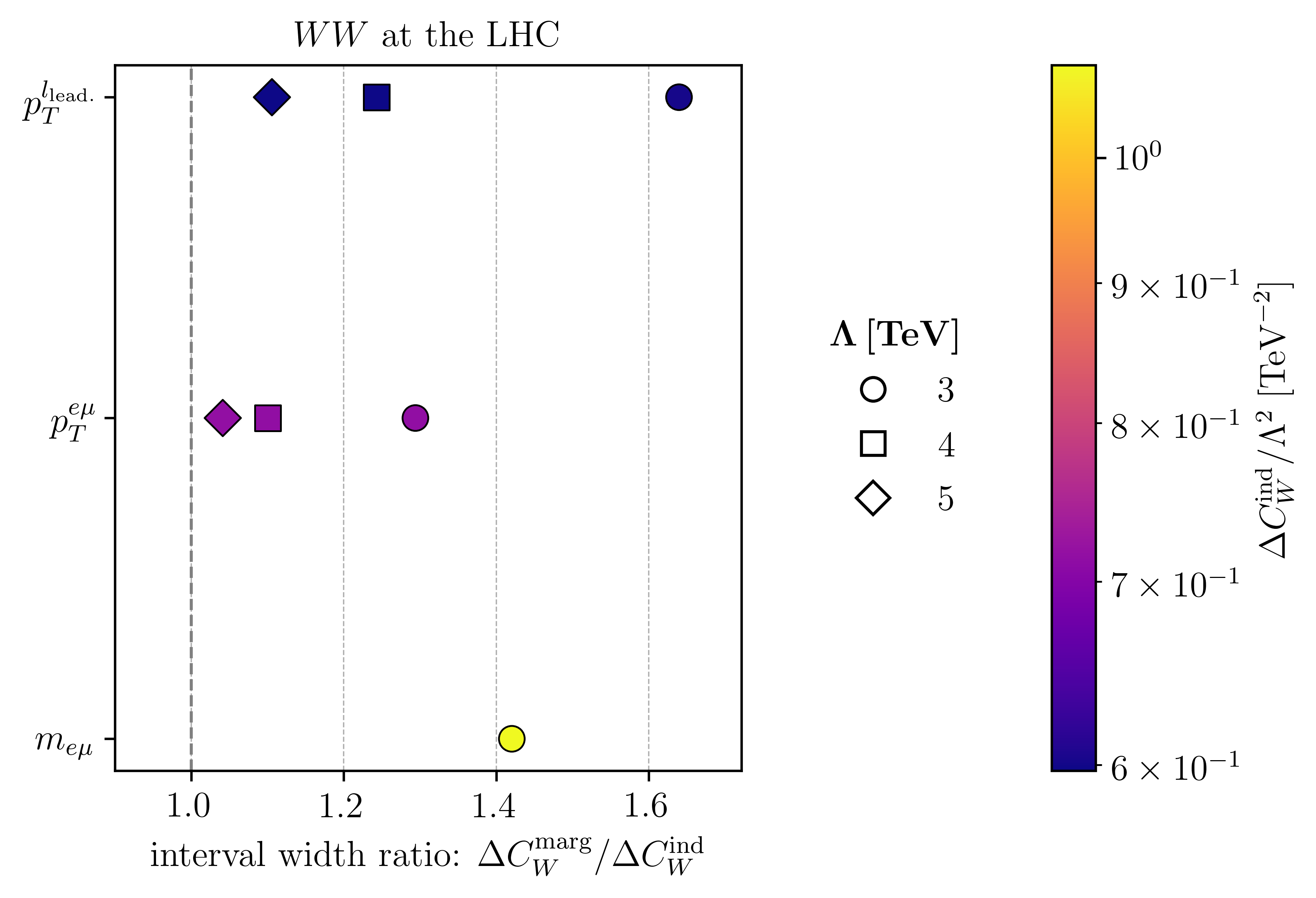}
    \caption{Same as~\cref{fig:wz_bounds_ratio} but for the $WW$ channel.}
    \label{fig:ww_bounds_ratio}
\end{figure}

\begin{figure}[ht]
    \centering
    \includegraphics[width=0.7\linewidth]{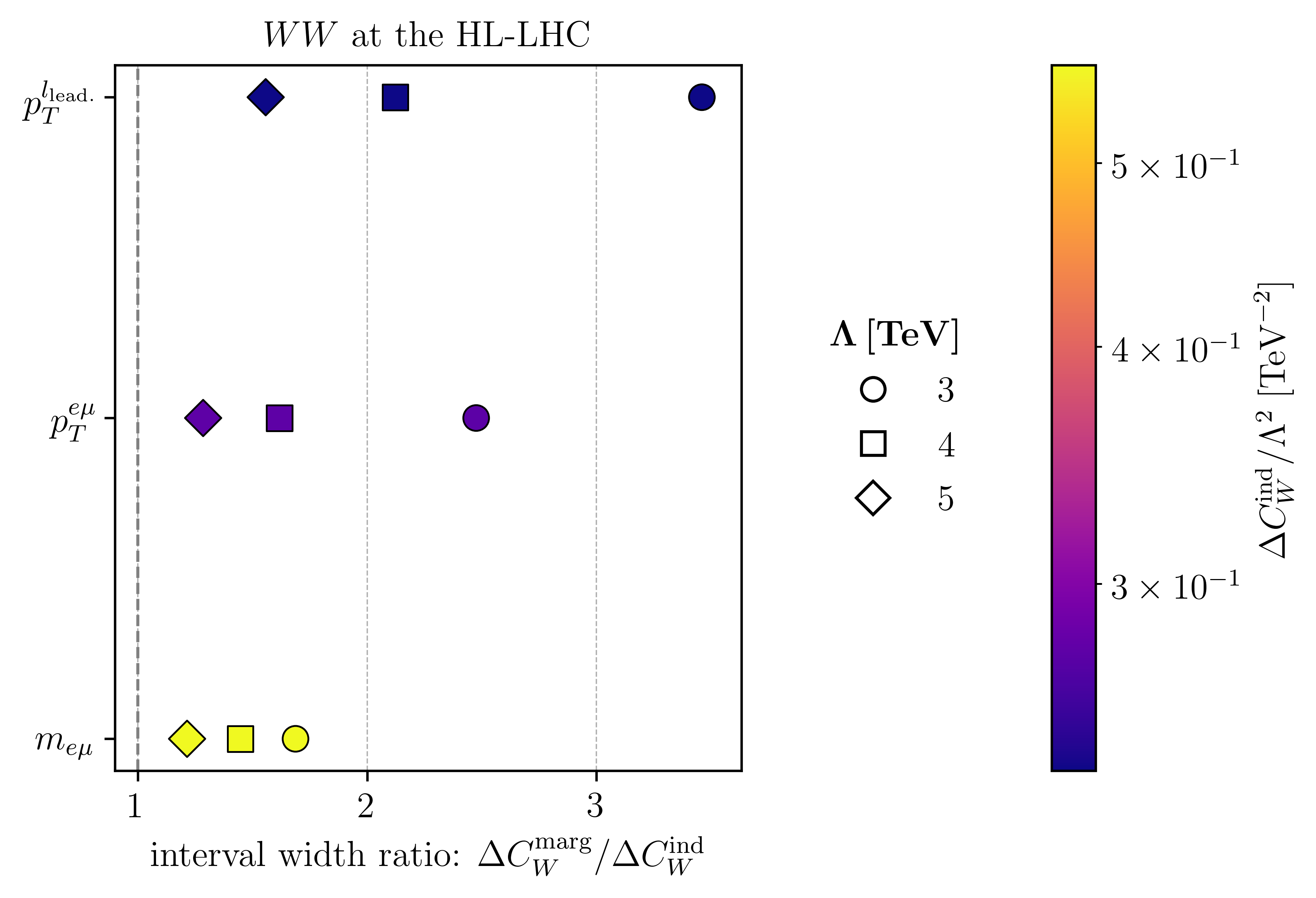}
    \caption{Same as~\cref{fig:ww_bounds_ratio} but for the HL-LHC projections.}
    \label{fig:ww_bounds_ratio_HL}
\end{figure}

To illustrate more clearly the impact of including dimension-8 operators in the fit, we present the ratio of the widths of the marginalised to the individual bounds on the dimension-6 coefficient \( C_W \). The marginalised bounds on $C_{W}$ at various cut-off-scales are presented in~\cref{tab:wz_cw_lambda_bounds_marg,tab:ww_cw_lambda_bounds_marg} of~\cref{app:add_results}. The results are shown, respectively, for \( WZ \) and \( WW \) production with current LHC data in~\cref{fig:wz_bounds_ratio,fig:ww_bounds_ratio}, and for the HL-LHC projections in~\cref{fig:wz_bounds_ratio_HL,fig:ww_bounds_ratio_HL}. Each marker on the plot indicates the used new-physics scale $\Lambda$ ranging from $3$ to $5~\text{TeV}$. Results are displayed for each observable. The heat-map indicates the stringency of the individual bounds. 

For a given observable, the overlap of the markers at a ratio close to unity implies that, for such observables, bounds on the dimension-6 coefficients can be reliably extracted independently of whether or not the relevant dimension-8 operators are included. In~\cref{fig:wz_bounds_ratio,fig:wz_bounds_ratio_HL}, we observe that, in the \( WZ \) channel, both with the current LHC data and with the HL-LHC projections, in a fit that does not include dimension-8 operators, the observables \( \Delta\Phi_{W,Z} \), \( \Delta y_{Z,\ell_{W}} \) and \( \cos \theta^{*}_{e} \) can be reliably used to extract bounds on \( C_{W} \). It is worth noting that only the ratios at \( 3~\text{TeV} \) are shown for the aforementioned angular observables. At higher cut-off scales, within the allowed scanning range of the fit, the coefficient \( C_{W} \) remains unconstrained. 

The situation is different for dimensionful observables: extracting bounds, at \( 3~\text{TeV} \), is not robust under the inclusion of dimension-8 operators, since the ratio of marginalised to individual bounds can induce a $\sim$ 15\% change for $p_{T}^{Z}$ and around $\sim$ 30\% for $m_{T}^{WZ}$ for the LHC case. We note here that the transverse mass  generally yields the most stringent bound among all observables. At higher cut-off scales, i.e. at \( 5~\text{TeV} \), these observables are expected to provide more reliable constraints in a dimension-6-only analysis.
These conclusions remain qualitatively the same in the HL-LHC case.

Moving to the $WW$ channel in~\cref{fig:ww_bounds_ratio,fig:ww_bounds_ratio_HL}, for the current LHC data and the HL-LHC projections, respectively, 
it is evident that the impact of dimension-8 operators is more significant relative to the $WZ$ case. This is not surprising given the importance of $\mathcal{O}_{10}$. Considering the most stringent observable $p_{T}^{l_{\rm{lead.}}}$ and a cut-off scale of 3 TeV, neglecting dimension-8 effects can induce uncertainties as large as $60\%$ in the LHC case. 

Furthermore, we observe that the dilepton invariant mass is not a reliable observable for the extraction of bounds in the LHC case. 
Using this observable with the current LHC data, the coefficient \( C_W \) remains unconstrained beyond \( 3~\text{TeV} \), and the only available constraint carries an error of nearly \( 40\% \) in a dimension-6-only analysis—as seen in~\cref{fig:ww_bounds_ratio}. In the HL-LHC scenario in~\cref{fig:ww_bounds_ratio_HL}, by contrast, the coefficient is constrained at higher cut-off scales beyond \( 3~\text{TeV} \).

Finally, it is worth noting that in the HL-LHC projections, the ratios of marginalised to individual bounds are generally larger than in the LHC case. This suggests that the impact of dimension-8 operators will remain non-negligible at higher luminosities, 
and may even become more pronounced than at the LHC.

\subsection{Two-dimensional scans}
\begin{figure}[ht]
    \centering
    \includegraphics[width=1.0\textwidth]{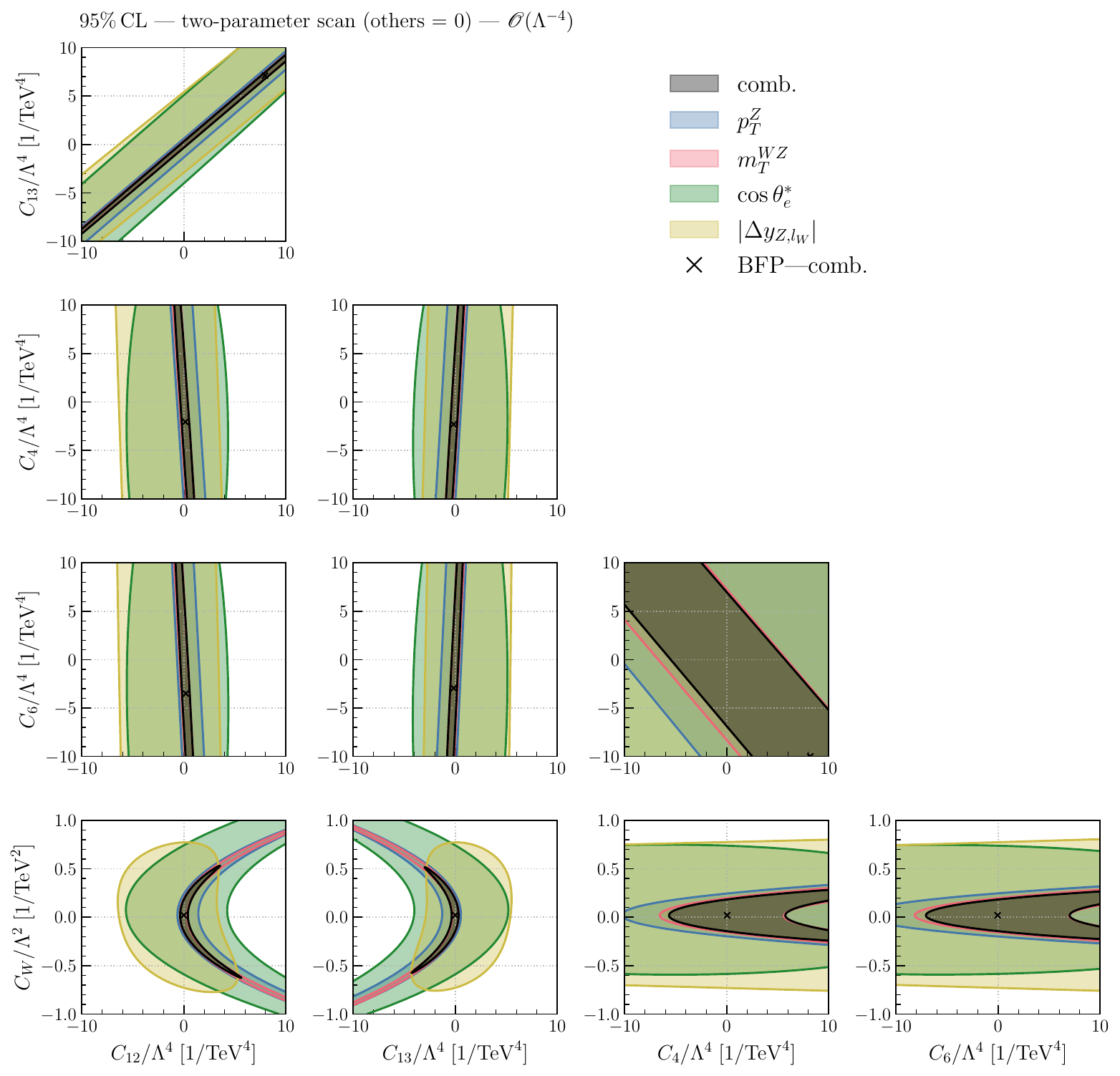}
    \caption{Two-dimensional fits for the CP-even dimension-6 and dimension-8 coefficients in the $WZ$ channel. Shown are the constraints from a selected set of observables individually, as well as from their combination. The EFT parametrisation includes linear contributions in dimension-8 as well as linear and quadratic contributions in dimension-6, i.e.~$\mathscr{O}(\Lambda^{-4})$. The best-fit point (BFP) of the combined fit is indicated.}
    \label{fig:wz_corner}
\end{figure}

\begin{figure}[ht]
    \centering
    \includegraphics[width=0.85\textwidth]{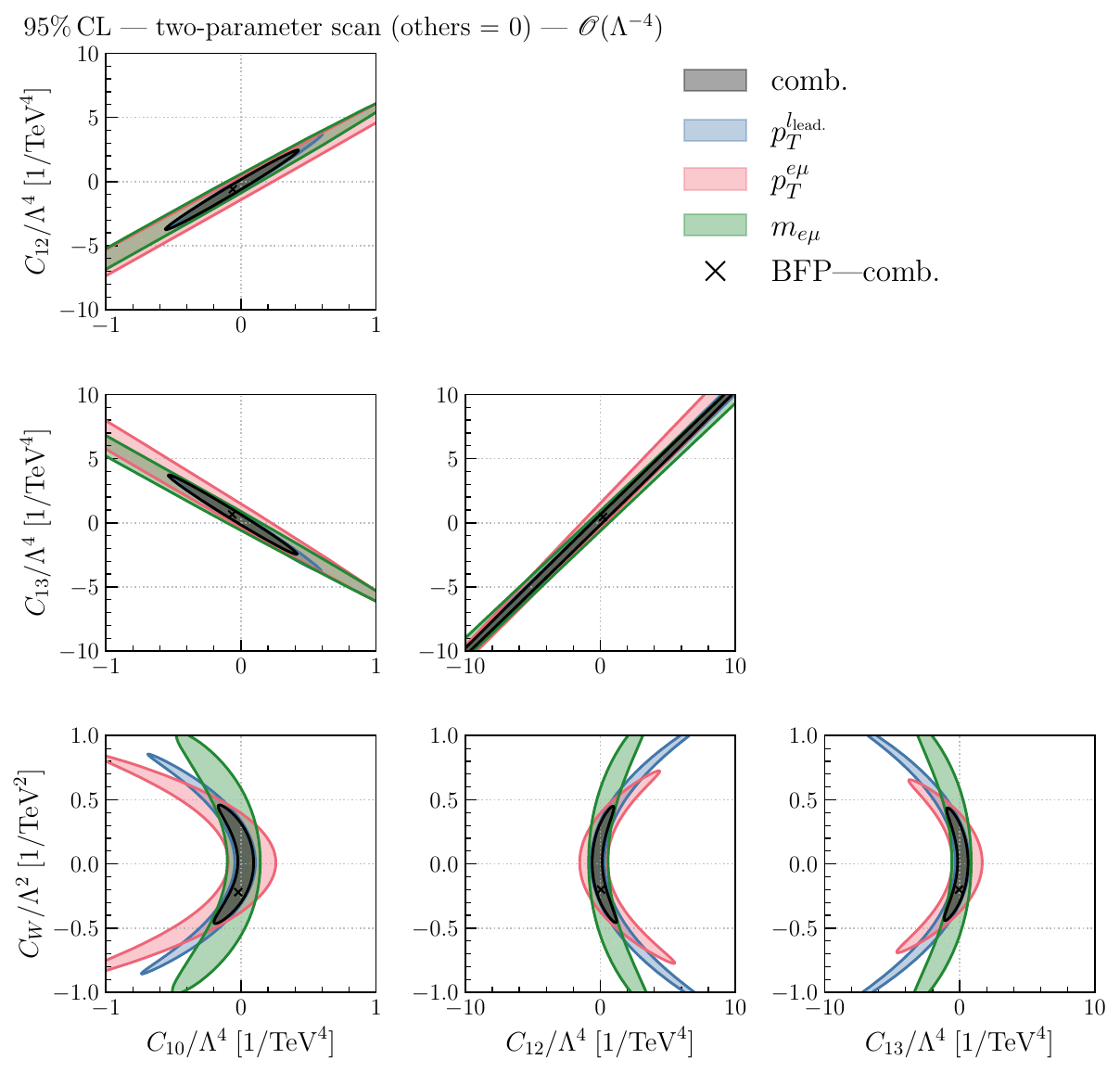}
    \caption{Same as~\cref{fig:wz_corner} but for the $WW$ process.}
    \label{fig:ww_corner}
\end{figure}

In this section, we aim to assess possible degeneracies between the operators we consider and thus we present the two-dimensional fits for the $WZ$ and $WW$ channels in~\cref{fig:wz_corner,fig:ww_corner}, respectively. The fit retains terms up to $\mathscr{O}(\Lambda^{-4})$ in the EFT expansion, and includes all the relevant dimension-6 and dimension-8 CP-even coefficients. The joint-probability scans include only a selected subset of observables for $WZ$, together with the contour corresponding to their combination. We emphasise that the latter is a simplified combination, assuming no bin-to-bin or cross-distribution correlations.

For the $WZ$ case shown in~\cref{fig:wz_corner}, one can immediately identify flat directions arising between the pairs ($C_{4}$, $C_{6}$) and ($C_{12}$, $C_{13}$). Across the board, $m_{T}^{WZ}$ is the observable with the strongest constraining power. An interesting feature appears in the scan of dimension-6 coefficient, $C_W$, vs the dimension-8 ones, $C_{12}$ and $C_{13}$: the contour arising from $\Delta y_{Z,l_{W}}$ exhibits a markedly different shape compared to the other observables. Since these joint probability scans essentially probe the shape dependence of the observables, we attribute this behaviour to the corresponding distribution shown in the right panel of~\cref{fig:wz_fid_cos_dy}. There, the rapidity separation at dimension-8 displays a distinctly different shape compared to both dimension-6 and the SM predictions. This feature is not observed, for instance, for the operators $\mathcal{O}_{4}$ and $\mathcal{O}_{6}$ where observe a correlation between either of $C_4$ or $C_6$ and the dimension-6 coefficient \(C_{W}\) in the case of \(\Delta y_{Z,\ell_{W}}\).

Finally, for the \(\cos \theta_{e}^{*}\) observable, we observe that, unlike the other cases, the corresponding contour exhibits a broken correlation between \(C_{W}\) and the 
dimension-8 coefficients $C_4$ and $C_{6}$, as well as between \(C_{4,6}\) and \(C_{12,13}\). This behaviour can be understood from the left panel of \cref{fig:wz_fid_cos_dy}, where the distinctive shape of each contribution is clearly visible and manifests in the fit.

In~\cref{fig:ww_corner} for the $WW$ channel, a correlation is again observed in the ($C_{12}$, $C_{13}$) pair across all observables. Furthermore, for each of their scan with $C_{10}$,  the transverse momentum of the leading lepton is observed to have the largest constraining power, in line with our findings in~\cref{tab:ww_all_bounds}.
In the dimension-6 scan, the strong constraining power of $C_{10}$ is once more manifest, with all observables competing to provide the most 
stringent bounds. For the dimension-6 scan against the pair ($C_{12}$, $C_{13}$), the contours are mirrored, as expected from the behaviour of these operators in the differential predictions, see for example,~\cref{fig:ww_fid_fully_incl_mass}. 

\section{Conclusions}
\label{sec:conclusions}
In this work, we studied diboson production, namely the $WZ$ and $WW$ processes, in the SMEFT framework, including the relevant dimension-6 and dimension-8 CP-even and CP-odd operators. We first classified the dimension-8 operators according to their high-energy behaviour and identified the subset that induces non-negligible effects. In doing so, we significantly reduced the number of dimension-8 operators entering the analysis and established that CP-odd effects remain negligible throughout.

We then computed SMEFT predictions at both the inclusive and differential cross-section level for all relevant operators in both channels, using a set of observables commonly employed in experimental analyses. Several noteworthy features were observed. In the $WZ$ channel,  the linear dimension-8 terms are subdominant compared to the squared dimension-6 ones for $\mathcal{O}(1)$ coefficients and $\Lambda=1$~TeV. In contrast, in the $WW$ channel, the contributions of the dimension-8 operator $\mathcal{O}_{10}$ are competitive with those from the dimension-6 operator. 

In addition, we computed the angular coefficients directly related to the longitudinal and transverse polarisation fractions,  and thereby quantifying the impact of dimension-8 operators on the vector-boson polarisation. We found negligible effects for a fully inclusive setup and mild, but non-negligible effects, for the production of transverse polarised bosons in the region $p_T^V > 400$ GeV. Furthermore, certain degeneracies between pairs of dimension-8 coefficients, previously identified in the differential distributions, were also reflected in this study.

Furthermore, we employed a $\chi^2$ fit to assess the impact of the different operators in constraining the corresponding coefficients. Using data from the ATLAS and CMS experiments, we compared our theoretical predictions at both the inclusive and differential level, performing both individual and marginalised fits. Our fit corroborates the earlier findings, namely that $C_{10}$ is relatively competitive in the $WW$ channel. We then assessed the effect of including or omitting dimension-8 operators in the fit. In this exercise, the marginalised bounds on the dimension-6 coefficient $C_{W}$ differed non-negligibly from the corresponding individual ones even for new physics scales of 3 TeV, underscoring the importance of including dimension-8 operators in a global fit scenario in a UV agnostic setup. We provided summary plots that encode which observables can be used reliably to extract bounds in analyses neglecting dimension-8 effects, and, in contrast, which observables require the inclusion of dimension-8 contributions in a marginalised fit. 

Finally, we presented two-dimensional scans for pairs of coefficients: for a selected set of observables in the $WZ$ channel and for all observables in the $WW$ channel. We also showed the contours corresponding to the combination of these observables. These results again confirmed our earlier findings, namely the presence of flat directions between the pairs of dimension-8 coefficients ($C_{4}, C_{6}$) and ($C_{12}, C_{13}$). In the $WZ$ channel, $m_{T}^{WZ}$ provides the strongest constraining power, whilst the rapidity separation $|\Delta y_{Z,l_{W}}|$ yields a distinctly shaped contour compared to the other observables. 
This observable can thus be used to disentangle dimension-6 and dimension-8 effects.   
In the $WW$ channel, the contours confirm the large impact of $O_{10}$.

We close by emphasising two important points. The first is the necessity of carrying out these studies at higher orders in perturbation theory. This is particularly relevant for the $WW$ channel, where jet-veto effects are significant and resummation effects cannot be neglected. As a first step, NLO QCD corrections to the dimension-8 SMEFT contributions should be computed to validate the robustness of our findings. Secondly, we stress that the aim of our study is to examine dimension-8 effects whilst remaining agnostic about the UV origin of the operators we consider. Using particular UV models which can then be matched to the EFT is necessary to establish the hierarchy among the Wilson coefficients and the relative size of dimension-6 and dimension-8 effects.  

\section*{Acknowledgements}
The authors are grateful to Hao-Lin Li for providing the FeynRules implementation of the model in Ref.~\cite{Degrande:2023iob}. HF thanks Víctor Miralles for useful input on the fit procedure. We also thank Eugenia Celada, Gauthier Durieux, Ken Mimasu, Alejo Rossia, and Marion Thomas for helpful discussions. This work was supported by the European Research Council (ERC) under the 
European Union’s Horizon 2020 research and innovation programme (Grant agreement No.~949451) and by a Royal Society University Research Fellowship through grant URF/R1/201553. HF acknowledges support from the European Union under the MSCA fellowship (Grant agreement No.~101208909).

\appendix
\section{Additional results}
\label{app:add_results}
In this section we present both marginalised and individual 95\%~CL bounds on the Wilson coefficient $C_{W}$ at various cut-off scales $\Lambda$. In Tables~\ref{tab:wz_cw_lambda_bounds_marg} and \ref{tab:wz_cw_lambda_bounds_ind} we show the results for $WZ$ production. In Tables~\ref{tab:ww_cw_lambda_bounds_marg} and \ref{tab:ww_cw_lambda_bounds_ind} we show the same for $WW$ production.
\begin{table}[ht]
    \renewcommand{\arraystretch}{1.25}
    \centering
    \begin{tabular}{l|c|c|c}
          \hline obs 
          & $C_W|_{|\Lambda=3 \rm{TeV}}$, marg  
          & $C_W|_{|\Lambda=4 \rm{TeV}}$, marg 
          & $C_W|_{|\Lambda=5 \rm{TeV}}$, marg 
          \\  \hline 
          $p_T^Z$ 
          & [-1.90, 2.30]
          & [-3.02, 3.73] 
          & [-4.57, 5.71] 
          \\ 
          $p_T^W$ 
          & [-2.11, 2.69]
          & [-3.33, 4.35] 
          & [-5.04, 6.59] 
          \\ 
          $m_T^{WZ}$ 
          & [-1.73, 2.09]
          & [-2.52, 3.19] 
          & [-3.69, 4.71] 
          \\
          $\Delta \Phi_{W,Z}$ 
          & [-6.45, 7.02]
          & –
          & –
          \\ 
          $p_T^{\nu}$ 
          & [-2.73, 3.40] 
          & [-4.61, 5.80]
          & [-7.11, 9.00] 
          \\ 
          $|\Delta y_{Z,l_{W}}|$ 
          & [-6.04, 6.28] 
          & –
          & – 
          \\ 
          $\cos \theta_e^*$ 
          & [-4.88, 6.26] 
          & –
          & –
          \\ \hline \hline
          $p_T^Z$ 
          & [-1.71, 2.26]
          & [-2.50, 3.47]
          & [-3.66, 5.16] 
          \\ 
          $p_T^W$ 
          & [-1.57, 2.09]
          & [-2.19, 3.11] 
          & [-3.11, 4.59] 
          \\ 
          $m_T^{WZ}$ 
          & [-1.26, 1.57]
          & [-1.42, 1.97] 
          & [-1.76, 2.59]
          \\
          $\Delta \Phi_{W,Z}$ 
          & [-3.88, 4.23]
          & [-6.66, 7.28]
          & –
          \\ 
          $p_T^{\nu}$ 
          & [-2.47, 3.21] 
          & [-4.00, 5.26]
          & [-6.04, 8.04] 
          \\ 
          $|\Delta y_{Z,l_{W}}|$ 
          & [-3.52, 4.14] 
          & [-6.04, 7.33]
          & – 
          \\ 
          $\cos \theta_e^*$ 
          & [-3.21, 4.47] 
          & [-5.45, 7.71]
          & –
          \\ \hline
    \end{tabular}
    \caption{Marginalised 95\%~CL bounds on the CP-even dimension-6 Wilson coefficient $C_{W}$ from $WZ$ production, shown per observable at NLO QCD. The results are presented for several cut-off scales $\Lambda$ and are obtained by retaining terms up to $\mathscr{O}(\Lambda^{-4})$. The upper block corresponds to LHC results, whilst the lower block shows the HL-LHC projections.}
    \label{tab:wz_cw_lambda_bounds_marg}
\end{table}
\begin{table}[ht]
    \renewcommand{\arraystretch}{1.25}
    \centering
    \begin{tabular}{l|c|c|c}
          \hline obs 
          & $C_W|_{|\Lambda=3 \rm{TeV}}$, ind 
          & $C_W|_{|\Lambda=4 \rm{TeV}}$, ind 
          & $C_W|_{|\Lambda=5 \rm{TeV}}$, ind
          \\  \hline 
          $p_T^Z$ 
          & [-1.61, 2.02]
          & [-2.87, 3.60]
          & [-4.48, 5.63] 
          \\ 
          $p_T^W$ 
          & [-1.76, 2.33]
          & [-3.14, 4.15] 
          & [-4.91, 6.48]
          \\ 
          $m_T^{WZ}$ 
          & [-1.26, 1.63]
          & [-2.25, 2.90] 
          & [-3.51, 4.54] 
          \\ 
          $\Delta \Phi_{W,Z}$ 
          & [-6.33, 6.89]
          & – 
          & –
          \\ 
          $p_T^{\nu}$ 
          & [-2.54, 3.21]
          & [-4.52, 5.71] 
          & [-7.06, 8.93]
          \\ 
          $|\Delta y_{Z,l_{W}}|$ 
          & [-6.00, 6.29]
          & –
          & – 
          \\ 
          $\cos \theta_e^*$ 
          & [-4.73, 6.13]
          & – 
          & –
          \\ \hline \hline
          $p_T^Z$ 
          & [-1.25, 1.79]
          & [-2.22, 3.19]
          & [-3.47, 4.98] 
          \\ 
          $p_T^W$ 
          & [-1.04, 1.57]
          & [-1.86, 2.80] 
          & [-2.90, 4.38] 
          \\ 
          $m_T^{WZ}$ 
          & [-0.49, 0.79]
          & [-0.87, 1.41] 
          & [-1.36, 2.20]
          \\
          $\Delta \Phi_{W,Z}$ 
          & [-3.69, 4.02]
          & [-6.57, 7.15]
          & –
          \\ 
          $p_T^{\nu}$ 
          & [-2.12, 2.84] 
          & [-3.78, 5.06]
          & [-5.90, 7.91] 
          \\ 
          $|\Delta y_{Z,l_{W}}|$ 
          & [-3.33, 4.11] 
          & [-5.93, 7.31]
          & – 
          \\ 
          $\cos \theta_e^*$ 
          & [-3.00, 4.27] 
          & [-5.34, 7.59]
          & –
          \\ \hline
    \end{tabular}
    \caption{Same as~\cref{tab:wz_cw_lambda_bounds_marg} but for individual bounds.}
    \label{tab:wz_cw_lambda_bounds_ind}
\end{table}
\begin{table}[ht]
    \renewcommand{\arraystretch}{1.25}
    \centering
    \begin{tabular}{l|c|c|c}
          \hline obs 
          & $C_W|_{|\Lambda=3 \rm{TeV}}$, marg  
          & $C_W|_{|\Lambda=4 \rm{TeV}}$, marg 
          & $C_W|_{|\Lambda=5 \rm{TeV}}$, marg 
          \\  \hline 
          $p_T^{{l}_{\rm lead.}}$ 
          & [-4.33, 4.54] 
          & [-5.76, 6.11] 
          & [-7.97, 8.54] 
          \\ 
          $p_T^{e\mu}$ 
          & [-4.09, 4.23] 
          & [-6.14, 6.45] 
          & [-9.09, 9.54] 
          \\ 
          $m_{e\mu}$ 
          & [-6.78, 7.04] 
          & – 
          & –
          \\ \hline \hline
          $p_T^{{l}_{\rm lead.}}$ 
          & [-3.59, 3.85] 
          & [-3.85, 4.28] 
          & [-4.33, 5.00] 
          \\ 
          $p_T^{e\mu}$ 
          & [-2.95, 3.19] 
          & [-3.35, 3.80] 
          & [-4.09, 4.80] 
          \\ 
          $m_{e\mu}$ 
          & [-4.09, 4.45] 
          & [-6.19, 6.85] 
          & [-8.02, 9.07] 
          \\ \hline
    \end{tabular}
    \caption{Same as~\cref{tab:wz_cw_lambda_bounds_marg} but for $WW$ production.}
    \label{tab:ww_cw_lambda_bounds_marg}
\end{table}
\begin{table}[ht]
    \renewcommand{\arraystretch}{1.25}
    \centering
    \begin{tabular}{l|c|c|c}
          \hline obs
          & $C_W|_{|\Lambda=3 \rm{TeV}}$, ind  
          & $C_W|_{|\Lambda=4 \rm{TeV}}$, ind 
          & $C_W|_{|\Lambda=5 \rm{TeV}}$, ind 
          \\  \hline 
          $p_T^{{l}_{\rm lead.}}$ 
          & [-2.60, 2.81] 
          & [-4.59, 4.96] 
          & [-7.17, 7.76] 
          \\ 
          $p_T^{e\mu}$
          & [-3.13, 3.30] 
          & [-5.57, 5.87] 
          & [-8.71, 9.18] 
          \\ 
           $m_{e\mu}$ 
          & [-4.73, 5.00] 
          & [-8.40, 8.89] 
          & – 
          \\ \hline \hline
          $p_T^{{l}_{\rm lead.}}$ 
          & [-0.95, 1.20] 
          & [-1.70, 2.13] 
          & [-2.66, 3.33] 
          \\ 
          $p_T^{e\mu}$
          & [-1.11, 1.37] 
          & [-1.98, 2.44] 
          & [-3.10, 3.82]  
          \\ 
           $m_{e\mu}$ 
          & [-2.34, 2.72] 
          & [-4.17, 4.84] 
          & [-6.51, 7.56] 
          \\ \hline
    \end{tabular}
    \caption{Same as~\cref{tab:ww_cw_lambda_bounds_marg} for individual bounds.}
    \label{tab:ww_cw_lambda_bounds_ind}
\end{table}

\bibliography{bibliography}
\bibliographystyle{JHEP}
\end{document}